\documentclass[conference,10pt]{IEEEtran}

\pdfoutput=1
\usepackage{cite}
\usepackage{authblk}
\usepackage{caption}
\newcommand\blfootnote[1]{%
  \begingroup
  \renewcommand\thefootnote{}\footnote{#1}%
  \addtocounter{footnote}{-1}%
  \endgroup
}

\usepackage{amsfonts}
\usepackage{amsmath}
\usepackage{amssymb}
\usepackage{color}
\usepackage[dvipsnames]{xcolor}
\usepackage{colortbl}
\definecolor{darkblue}{rgb}{0.1,0.2,0.6}
\definecolor{darkred}{rgb}{0.8,0.1,0.2}
\definecolor{Gray}{gray}{0.9}
\usepackage[colorlinks,citecolor=darkblue,linkcolor=darkred,urlcolor=darkblue]{hyperref}	
\usepackage[all]{hypcap} 

\usepackage{graphicx}
\graphicspath{{images/}}

\usepackage{enumitem}
\usepackage{listings}
\usepackage[group-separator={,}]{siunitx}

\newenvironment{allintypewriter}{\ttfamily}{\par}

\usepackage{blindtext}

\usepackage{xifthen}
\newboolean{comments}

\setboolean{comments}{true}

\ifthenelse{\boolean{comments}}
{\usepackage[markup=underlined]{changes}}
{\usepackage[final, markup=underlined]{changes}}

\definechangesauthor[color=OrangeRed]{TSH}
\definechangesauthor[color=BurntOrange]{SM}
\definechangesauthor[color=Purple]{SB}
\definechangesauthor[color=ForestGreen]{BV}
\definechangesauthor[color=Blue]{CH}
\definechangesauthor[color=Brown]{EGR}
\definechangesauthor[color=Red]{DL}

\ifthenelse{\boolean{comments}}
{

\newcommand{\tsh}[1]{\textcolor{OrangeRed}{\textsf{[\textbf{TSH:} #1]}}}
\newcommand{\sm}[1]{\textcolor{BurntOrange}{\textsf{[\textbf{SM:} #1]}}}
\newcommand{\sbx}[1]{\textcolor{Purple}{\textsf{[\textbf{SB:} #1]}}}
\newcommand{\bv}[1]{\textcolor{ForestGreen}{\textsf{[\textbf{BV:} #1]}}}
\newcommand{\ch}[1]{\textcolor{Blue}{\textsf{[\textbf{CH:} #1]}}}
\newcommand{\egr}[1]{\textcolor{Brown}{\textsf{[\textbf{EGR:} #1]}}}
\newcommand{\dl}[1]{\textcolor{Red}{\textsf{[\textbf{DL:} #1]}}}
}
{

\newcommand{\tsh}[1]{}
\newcommand{\sm}[1]{}
\newcommand{\sbx}[1]{}
\newcommand{\bv}[1]{}
\newcommand{\ch}[1]{}
\newcommand{\egr}[1]{}
\newcommand{\dl}[1]{}
}

\def\urbana{
	Institute for Condensed Matter Theory and Department of Physics, 
	University of Illinois at Urbana-Champaign, Urbana, IL 61801, USA
}

\def\quail{
	Quantum Artificial Intelligence Lab. (QuAIL), 
    NASA Ames Research Center, Moffett Field, CA 94035, USA
}

\def\usra{
	USRA Research Institute for Advanced Computer Science (RIACS),
	615 National, Mountain View, California 94043, USA
}

\def\google{
	Google Inc., Venice, CA 90291, USA
}

\def\sgt{
	Stinger Ghaffarian Technologies Inc., 
    7701 Greenbelt Rd., Suite 400, Greenbelt, MD 20770, USA
}

\def\ornl{
    Quantum Computing Institute, Oak Ridge National Laboratory, 
    Oak Ridge, TN, 37831, USA
}

\def\olcf{
    Scientific Computing, Oak Ridge Leadership Computing, Oak Ridge National Laboratory, 
    Oak Ridge, TN, 37831, USA
}

\begin{document}

\title{Establishing the Quantum Supremacy Frontier with a 281 Pflop/s Simulation\vspace{-20pt}}

\author[1,2,3,$*$]{Benjamin Villalonga}
\author[4,5$\dagger$]{Dmitry Lyakh}
\author[6,$\ddagger$]{Sergio Boixo}
\author[6,$+$]{Hartmut Neven}
\author[1,$\times$]{Travis S.~Humble$^{4,\mathsection}$,\\Rupak Biswas}
\author[1,$\mathparagraph$]{Eleanor G. Rieffel}
\author[6,$\|$]{Alan Ho}
\author[1,7,$\vdash$]{Salvatore Mandr\`a}
\affil[1]{\small\quail}
\affil[2]{\usra}
\affil[3]{\urbana}
\affil[4]{\ornl}
\affil[5]{\olcf}
\affil[6]{\google}
\affil[7]{\sgt}

\maketitle

\begin{abstract}
   Noisy Intermediate-Scale Quantum (NISQ) computers are  entering an era in
   which they can perform computational tasks beyond the capabilities of the
   most powerful classical computers, thereby achieving ``Quantum Supremacy'', a
   major milestone in quantum computing. NISQ Supremacy requires comparison with
   a state-of-the-art classical simulator.  We report HPC simulations of hard
   random quantum circuits (RQC), which have been recently used as a benchmark
   for the first experimental demonstration of Quantum Supremacy, sustaining an
   average performance of 281 Pflop/s (true single precision) on Summit,
   currently the fastest supercomputer in the World.  These simulations were
   carried out using qFlex, a tensor-network-based classical high-performance
   simulator of RQCs.  Our results show an advantage of many orders of magnitude
   in energy consumption of NISQ devices over classical supercomputers.  In
   addition, we propose a standard benchmark for NISQ computers based on qFlex.
\end{abstract}

\blfootnote{%
\hspace{-10pt}\begin{tabular}{l}
  $*$ \href{mailto:vlllngc2@illinois.edu}{vlllngc2@illinois.edu}, 
  $\dagger$ \href{mailto:liakhdi@ornl.gov}{liakhdi@ornl.gov},
  $\ddagger$ \href{mailto:boixo@google.com}{boixo@google.com},\\
  $+$ \href{mailto:neven@google.com}{neven@google.com},
  $\mathsection$ \href{mailto:humblets@ornl.gov}{humblets@ornl.gov}, 
  $\times$ \href{mailto:rupak.biswas@nasa.gov}{rupak.biswas@nasa.gov},\\
  $\mathparagraph$ \href{mailto:eleanor.rieffel@nasa.gov}{eleanor.rieffel@nasa.gov},
  $\|$ \href{mailto:alankho@google.com}{alankho@google.com},\\
  $\vdash$ \href{mailto:salvatore.mandra@nasa.gov}{salvatore.mandra@nasa.gov}
\end{tabular}
}\vspace{-15pt}

\section{Introduction and Motivations}

As we approach the end of Moore's law, there is growing urgency to develop
alternative computational models. Examples are Beyond Von Neumann architectures
and Beyond CMOS hardware. Since Richard Feynman envisioned the possibility of
using quantum physics for computation \cite{feynman1982simulating,
feynman1985quantum}, many theoretical and technological advances have been made.
Quantum computing is both Beyond Von Neumann and Beyond CMOS. It represents the
most potentially transformative technology in computing. 

Recent experiments have underscored the remarkable potential for quantum
computing to offer new capabilities for computation. Experiments using
superconducting electronic circuits have shown that control over quantum
mechanics can be scaled to modest sizes with relatively good accuracy
\cite{barends_superconducting_2014, kelly_state_2015, wang201816,
barends2016digitized, arute2019quantum}.
By exercising control over quantum physical systems, quantum computers are
expected to accelerate the time to solution and/or accuracy of a variety of
different applications~\cite{NCbook, RPbook, grover1996fast,
feynman1982simulating, aspuru-guzik_simulated_2005, babbush_low_2017,
jiang_quantum_2018, babbush_encoding_2018}, while simultaneously reducing power
consumption. 
For instance, important encryption protocols such as RSA, which is widely used
in private and secure communications, can theoretically be broken using quantum
computers \cite{shor1994algorithms}. 
Moreover, quantum simulation problems, including new material design and drug
discovery, could be accelerated using quantum processors
\cite{feynman1982simulating, aspuru-guzik_simulated_2005, babbush_low_2017,
jiang_quantum_2018, babbush_encoding_2018,smelyanskiy2018non}.\\

However, building a universal, fault tolerant quantum computer is, at the
moment, a long-term goal driven by strong theoretical \cite{fowler2012surface,
boixo_characterizing_2018, bremner_achieving_2017} and experimental
\cite{barends_superconducting_2014, kelly_state_2015, neill_blueprint_2018}
evidence. 
Nonetheless, despite the lack of error correction mechanisms to run arbitrary
quantum algorithms, Noisy Intermediate-Scale Quantum (NISQ) devices of about
50-100 qubits are expected to perform certain tasks which surpass the
capabilities of classical high-performance computer systems
\cite{de_raedt_massively_2007, preskill_quantum_2018, bremner_achieving_2017,
smelyanskiy_qhipster:_2016, boixo_characterizing_2018, haner20170,
aaronson2016complexity, bouland_quantum_2018, pednault_breaking_2017,
chen_64-qubit_2018, li_quantum_2018, neill_blueprint_2018,
chen_classical_2018,markov_quantum_2018, villalonga2018flexible,
de_raedt_massively_2018}, therefore achieving Quantum Supremacy.
Examples of NISQ devices are the 53-qubit IBM Rochester QPU~\cite{ibm_2019}, and
the 53-qubit Google Sycamore QPU~\cite{arute2019quantum}. Both NISQ devices are
``digital'' universal quantum computers, namely they perform a universal set of
discrete operations (``gates'') on qubits. Quantum algorithms are translated
into quantum circuits and run on the NISQ device. The ``depth'' of quantum
circuits is defined as the number of clock cycles, where each clock cycle is
composed of gate operations executed on distinct qubits. Given the noisy nature
of NISQ devices, only shallow or low depth quantum circuits can be run.
 
An extensive body of work in complexity theory in the context of "Quantum
Supremacy" and random circuit sampling (RCS) supports the conclusion that ideal
quantum computers can break the Strong Church-Turing thesis by performing
computations that cannot be reduced in polynomial time to any classical computational
model~\cite{aaronson2011computational, bremner_average-case_2016,
bremner_achieving_2017, boixo_characterizing_2018, aaronson2017complexity,
bouland_quantum_2018,harrow2018approximate, movassagh_efficient_2018}.  The
milestone of experimentally demonstrating Quantum Supremacy was recently
achieved by Google with its Sycamore processor~\cite{arute2019quantum}.

Here we present a state-of-the-art classical high-performance simulator of large
random quantum circuits (RQCs), called qFlex, and propose a systematic benchmark
for Noisy Intermediate-Scale Quantum (NISQ)
devices~\cite{preskill_quantum_2018}. qFlex makes use of efficient
single-precision matrix-matrix multiplications and an optimized tensor transpose
kernel on NVIDIA GPUs. It utilizes efficient asynchronous pipelined task-based
execution of tensor contractions implemented in its portable computational
backend (TAL-SH library) that can run on both multicore CPU and NVIDIA GPU
architectures.  Our GPU implementation of qFlex reaches a peak performance
efficiency of 92\% and delivers a sustained absolute performance efficiency of
68\% in simulations spanning the entire Summit supercomputer (281 Pflop/s single
precision without mixed-precision acceleration). Due to our
communication-avoiding algorithm, this performance is stable with respect to the
number of utilized nodes, thus also demonstrating excellent strong scaling.

\subsection{RCS protocol}

While these demonstrations show progress in the development of quantum
computing, it remains challenging to compare alternative quantum computing
platforms and to evaluate advances with respect to the current dominant
(classical) computing paradigm. The leading proposal to fulfill this critical
need is quantum random circuit sampling (RCS) \cite{aaronson2017complexity,
bouland_quantum_2018, movassagh_efficient_2018, boixo_characterizing_2018,
arute2019quantum}, that is, to approximately sample bitstrings from the output
distribution generated by a random quantum circuit.  This is a "hello world"
program for quantum computers, because it is the task they are designed to do.
This proposal also counts with the strongest theoretical support in complexity
theory for an exponential separation between classical and quantum computation
\cite{aaronson2017complexity, bremner_achieving_2017}.  It is precisely this
``hello world'' program that was used in Ref.~\cite{arute2019quantum} to
demonstrate Quantum Supremacy.  Classically, it requires a simulation of the
quantum computation, and the computational cost increases exponentially in the
number of qubits for quantum circuits of sufficient depth.  Storing the wave
function from the output of a random quantum circuit with 40 qubits requires 8.8
TB at single precision ($2^{40} \times 8$ bytes), and 50 qubits require 9 PB.
Although new algorithms such as qFlex are designed to avoid these memory
requirements, calculating the necessary large number of probabilities for the
output bitstrings of quantum circuits of sufficient depth with 40 qubits or more
can only be achieved with
HPC resources.

The RCS protocol to benchmark NISQ computers consists of the following steps:
\begin{enumerate}
\item Generate random circuits fitted to a given NISQ architecture,
\item Execute random circuits, collecting bitstrings, time to solution, and
      energy consumption of the quantum computer,
\item Calculate probabilities of collected bitstrings using a classical
      simulation (qFlex) to measure the NISQ performance or fidelity (see
      below),
\item Perform a classical noisy simulation (qFlex) based on calculated (point 3)
      fidelity to collect equivalent classical computational metrics of
      relevance -- total floating point operations/time-to-solution/power
      consumption.
\end{enumerate}

\begin{figure}
\fbox{\begin{minipage}{\columnwidth}
  \begin{allintypewriter}
  
    circuit = RandomCircuit()
    \vskip 0.1in
    size = 10**6
    \vskip 0.1in
    \# Run on real hardware.
    
    samples = Sample(circuit, size)
    \vskip 0.1in
    \# Determine probabilities with simulation
    
    p = 1
    
    for s in samples:
    
      \hskip .2 in 
      \# EXPENSIVE!
      
      \hskip .2 in
      p *= Probability(circuit, s)
    \vskip 0.1in
    \# Derive cross entropy.
    
    cross\_entropy = -log(p) / size
  
  \end{allintypewriter}
\end{minipage}}
\caption{\label{fig:xeb}Cross-entropy benchmarking (XEB) pseudo-code.}
\end{figure}

\noindent The NISQ fidelity (item 3 above) of the RCS protocol can be estimated
using qFlex to calculate the probabilities of the bitstrings obtained in the
NISQ experiment. This is done by using the cross-entropy benchmarking (XEB), see
Fig.~\ref{fig:xeb} and Ref.~\cite{boixo_characterizing_2018}. Then, the
so-called heavy output generation (HOG) score is also calculated with the same
data \cite{aaronson2017complexity}. Note that items 3 and 4 above cannot be
performed once the Quantum Supremacy threshold has been surpassed. In that case,
the fidelity of the quantum computer can only be estimated for easier instances
of the RQCs, an estimate that proves accurate in
practice~\cite{arute2019quantum}.
An alternative benchmark for NISQ devices called ``Quantum
Volume''~\cite{bishop2017quantum} is also based on RCS with additional
requirements in the model circuits (item 1 above).

\subsection{Motivation for RCS as a Standard Benchmark}

The purpose of benchmarking quantum computers is twofold: to
objectively rank NISQ computers as well as to understand
quantum computational power in terms of equivalent “classical computing”
capability. Therefore, to achieve the above objectives, a good benchmark for
NISQ computers must have the following properties:

\begin{itemize}[label={-}, leftmargin=\parindent]
  \item\textbf{A good predictor of NISQ algorithm performance.} In the classical
        computing world,  Linpack is used as a benchmark for HPC Top-500 and serves as a
        relatively good predictor of application performance. Similarly, RCS
        serves as a good “multi-qubit” benchmark because:
        \begin{itemize}[label={$\circ$}, leftmargin=20pt]
          \item The computation is well defined,
          \item The computation requires careful control of multiple qubits,
          \item It measures the fidelity of a NISQ computer using cross-entropy
                benchmarking (XEB) \cite{boixo_characterizing_2018,
                neill_blueprint_2018, arute2019quantum}
          \item Works for any set of universal gates.
        \end{itemize}\vspace{4pt}
  \item\textbf{Estimating the equivalent classical worst-case computation
        resources} with commonly used metrics such as time-to-solution speedup,
        equivalent floating point operations, and equivalent power usage
        metrics.  Typically, advancements in classical algorithms could
        invalidate a quantum benchmark result. A key feature of RCS is that
        there is a large body of theory in computational complexity against an
        efficient classical algorithm \cite{aaronson2011computational,
        bremner_average-case_2016, bremner_achieving_2017,
        boixo_characterizing_2018, aaronson2017complexity,
        bouland_quantum_2018,harrow2018approximate, movassagh_efficient_2018}.
        Furthermore, qFlex for RCS requires computational resources
        proportional to the sampling fidelity, taking into account NISQ errors
        \cite{markov_quantum_2018,villalonga2018flexible}. Lastly, quantum
        random circuits produce quasi-maximal entanglement in the ideal
        case.\vspace{4pt}
  \item\textbf{Architectural neutrality.} Different architectures (e.g. ion
        traps vs superconducting qubits) have very different performance
        capabilities in terms of:
        \begin{itemize}[label={$\circ$}, leftmargin=20pt]
          \item Number of qubits
          \item Types of gates and fidelity
          \item Connectivity
          \item Number of qubits per multi-qubit operation
          \item Number of parallel executions
        \end{itemize}
        RCS allows a fair comparison across architectures by estimating the
        classical computational power of each  given sampling task.\vspace{4pt}
  \item\textbf{Useful for calibration of NISQ devices.} Critical issues such as
        cross-talk \cite{barends_superconducting_2014, kelly_state_2015,
        neill_blueprint_2018, arute2019quantum} only come up when scaling
        multi-qubit computations. XEB with RCS allows for calibration on
        different sets of qubits at different depths.
\end{itemize}
Other techniques for benchmarking NISQ devices, such as randomized benchmarking
\cite{magesan_robust_2011,knill2008randomized,magesan_characterizing_2012},
cycle benchmarking \cite{erhard2019characterizing} or preparing entangled GHZ
states \cite{wang201816} are restricted to non-universal gate sets and easy to
simulate quantum circuits. This is an advantage for extracting circuit
fidelities of non-universal gate sets, but does not measure computational power. 

\subsection{Energy Advantages of Quantum Computing}
\label{sec:energy_advantages}

The largest casualty as we approach the end of Moore's law has been the energy
efficiency gains due to Dennard scaling, also known as Koomey's law. Today's HPC
data centers are usually built within the constraints of available energy
supplies, rather than the constraints in hardware costs. For example, the Summit
HPC system at Oak Ridge National Laboratory has a total power capacity of  14 MW
available to achieve a design specification of 200 Pflop/s double-precision
performance. To scale such a system by 10x would require 140 MW of power, which
would be prohibitively expensive. Quantum computers on the other hand have the
opportunity to drastically reduce power consumption \cite{Britt2017,
Humble2018}.

\begin{figure}
\centering
\includegraphics[width=128.8pt,trim={62pt 0pt 4pt 0pt},clip]{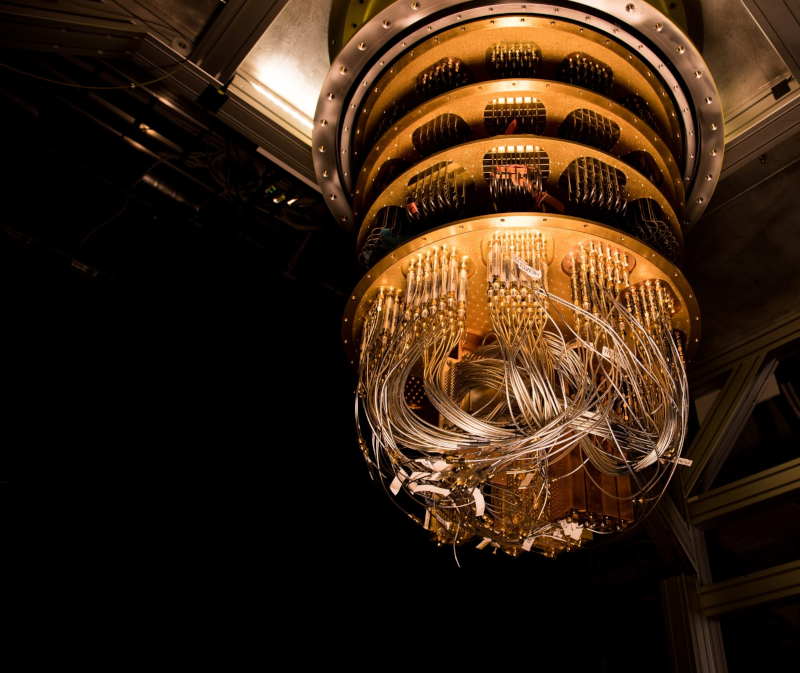}
\includegraphics[width=110pt]{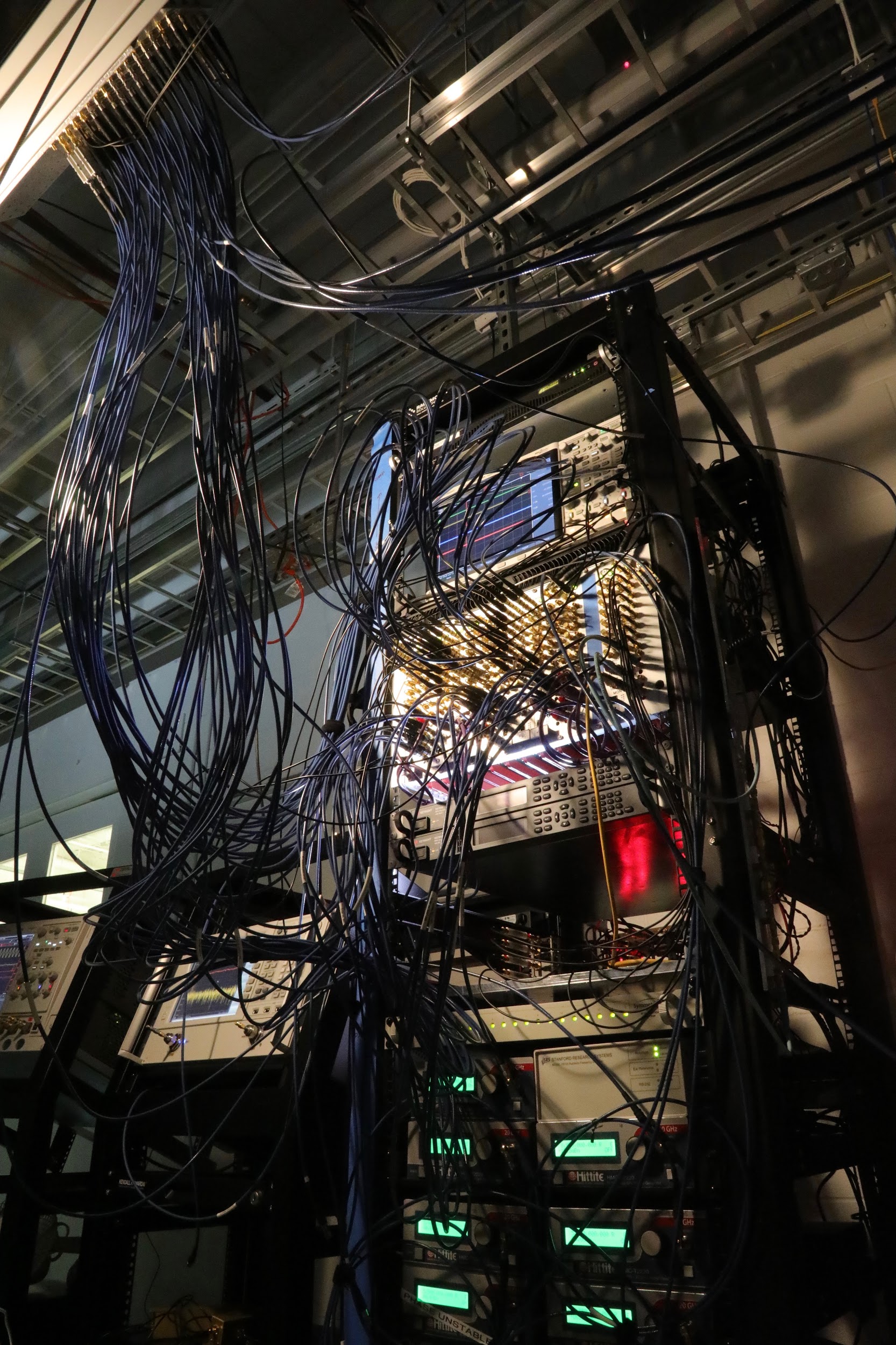}
\caption{\label{fig:refrigerator}A side by side view of a dilution refrigerator
and electronic racks for Google's quantum computer.}
\end{figure}

To compare the energy consumption needed for a classical computation to the
corresponding needs of a quantum computation, it is important to understand the
sources of energy consumption for quantum computers. For superconducting quantum
computers such as Google's (Fig.~\ref{fig:refrigerator}), the main sources of
energy consumption are:

\begin{enumerate}[label={-}, leftmargin=\parindent]
  \item Dilution Refrigerator - because quantum computers operate at around the
        15 mK range, a large dilution refrigerator that typically consumes $\sim$10
        kW is necessary.
  \item Electronic racks - these typically consume around 5 kW of power. They
        consist of oscilloscopes, microwave electronics, Analog-Digital
        converters, and clocks.
\end{enumerate}

\noindent Therefore a typical quantum computer will consume approximately
$\sim$15 kW of power for a single QPU of 53 qubits. Even as qubit systems scale
up, this amount is unlikely to significantly grow. The main reasons are:
dilution refrigerators are not expected to scale up; the power for the microwave
electronics on a per-qubit basis will decrease as quantum chips scale in the
number of qubits. The last component needed to perform a fair comparison between
classical and quantum computing is the rate at which quantum computers can
execute a circuit. In the case of superconducting qubits, the circuit execution
rate is approximately 10 kHz to 100 kHz.

For classical computers, different architectures can result in different energy
consumption rates. In this paper, we compare the consumption of NASA's
supercomputer Electra and ORNL's supercomputer Summit. On ORNL's Summit, each
node consumes around 2.4 kW of energy and consists of 2 IBM Power9 CPUs and 6
NVIDIA V100 GPUs, with a total of 4608 nodes (1.8 kW per node is consumed by 6
NVIDIA Volta GPUs). On NASA's Electra supercomputer, each of the 2304 40-core
Skylake nodes consumes about 0.52 kW, and each of the 1152 28-core Broadwell
nodes consumes about 0.38 kW.

\section{Current State-of-the-art}
\label{sec:soa}

With the advent of NISQ devices, different classes of classical simulators have
blossomed to tackle the demanding problem of simulating large RQCs. To date,
classical simulators of RQCs can be classified into 
three main classes (plus a fourth category of ``hybrid'' algorithms):\\

\begin{itemize}[label={-}, leftmargin=\parindent]
  \item\textbf{Direct evolution of the quantum state.} This approach consists in
        the sequential application of quantum gates to the full representation of an
        $n-$qubit state. While this method does not have any limitation in
        terms of gates and topology of RQCs, the memory required to store the
        full quantum state quickly explodes as $2^n$, and taking into account the
        non-negligible overhead given by the node-to-node communication to
        update the full quantum state. Thus, this approach becomes impractical for the
        simulation and benchmark of NISQ devices with $\sim 50$ qubits. Beyond
        $\sim$40 qubits it already requires fast internode connectivity
        \cite{de_raedt_massively_2007, smelyanskiy_qhipster:_2016, haner20170,
        pednault_breaking_2017, de_raedt_massively_2018}.\vspace{4pt}

  \item\textbf{Perturbation of stabilizer circuits.} It is known that stabilizer
        circuits, i.e.  quantum circuits made exclusively of Clifford gates, can
        be efficiently simulated and sampled on classical computers
        \cite{gottesman1998heisenberg}. An intriguing idea consists in
        representing RQCs as a linear combination of different stabilizer
        circuits \cite{aaronson2004improved, bravyi_improved_2016,bennink2017}.  Therefore,
        the computational complexity to simulate RQCs becomes proportional to
        the number $\chi$ of stabilizer circuits required to represent them.
        Unfortunately, $\chi$ doubles every time a non-Clifford gate (for
        instance, a $\tfrac{\pi}{8}$ rotation) is used.  Furthermore, the number
        of non-Clifford gates grows quickly with the depth of RQCs
        \cite{boixo_characterizing_2018} and, in general, much faster than the
        number of qubits.  Consequently, stabilizer circuits are not suitable
        to benchmark NISQ devices using RQCs.\vspace{4pt}

  \item\textbf{Tensor network contractions.} The main idea of this approach is
        based on the fact that any given quantum circuit can always be
        represented as a tensor network, where one-qubit gates are rank-2
        tensors (tensors of 2 indexes with dimension 2 each), two-qubit gates
        are rank-4 tensors (tensors of 4 indexes with dimension 2 each), and in
        general $n$-qubit gates are rank-$2n$ tensors
        \cite{markov_simulating_2008, biamonte2017tensor} (this approach should
        not be confused with another application of tensor network theory to
        quantum circuit simulations where tensor networks are used for
        wave-function compression \cite{TNQVM}). In general, the computational
        and memory cost for contracting such networks is (at least) exponential
        with the number of open indexes (corresponding to the input state and
        output state, respectively).  Therefore, for large enough circuits, the
        network contraction is impractical. Nonetheless, once the input and
        output states are specified through rank-1 Kronecker projectors, the
        computational complexity drastically reduces. More precisely, the
        computational and memory costs are dominated by the largest tensor during
        the contraction \cite{markov_simulating_2008}. The size of the largest
        contraction can be estimated by computing the treewidth of the
        underlying tensor network's line graph which is, for most of the NISQ
        devices, proportional to the depth of the RQCs.  This representation of
        quantum circuits gives rise to an efficient simulation technique to
        simulate RQCs for XEB.  Among the most relevant tensor-based simulators,
        it is worth to mention the following:\vspace{4pt}

        \begin{itemize}[label={$\circ$}]
          \item\textbf{Undirected graphical model.} An approach based on
                algorithms designed for undirected graphical models, closely
                related to tensor network contractions, was introduced in
                \cite{boixo_simulation_2017}. Later, the Alibaba group used the
                undirected graphical representation of RQCs to take advantage of
                underlying weaknesses of the design \cite{chen_classical_2018}.
                More precisely, by carefully projecting the state of a few nodes in the graph,
                the Alibaba group was able to reduce the computational
                complexity to simulate large RQCs.  However, it is important to
                stress that \cite{chen_classical_2018} reports the computational
                cost to simulate a class of RQCs which is much easier to
                simulate than the class of RQCs reported in
                Ref.~\cite{boixo_characterizing_2018}. Indeed, Chen et al. fail
                to include the final layer of Hadamard gates in their RQCs and
                use more ${\rm T}$ gates in detriment of non-diagonal gates at the beginning of the circuit. For
                these reasons, we estimate that such a circuit class is about $1000\times$
                easier to simulate than the RQCs available in
                \cite{new_benchmarks}, as discussed in Ref.~\cite{villalonga2018flexible}. The latter are the circuits simulated in
                this submission, as well as in \cite{markov_quantum_2018,
                villalonga2018flexible}. See
                Appendix~\ref{sec:variable_elimination} for more details about
                this simulator. \vspace{4pt}
                
          \item\textbf{Quantum teleportation-inspired algorithms.} Recently,
                Chen et al. have proposed a related approach for
                classical simulation of RQCs inspired by quantum teleportation
                \cite{chen_quantum_2019}.  Such approach allows to ``swap''
                space and time to take advantage of low-depth quantum circuits
                (equivalent to a Wick rotation in the imaginary time direction).
                While the approach is interesting \emph{per se}, the
                computational performance is lower than that one achieved by qFlex
                \cite{villalonga2018flexible}.
        \end{itemize}\vspace{4pt}

  \item\textbf{Hybrid algorithms.} Several works have explored algorithms based
        on splitting grids of qubits in smaller sub-circuits, which are then
        independently simulated \cite{chen_64-qubit_2018, markov_quantum_2018}.
        As an example, every time a ${\rm CZ}$-gate crosses between
        sub-circuits, the number of independent circuits to simulate is
        duplicated.
        Therefore, the computational cost is exponential in the number of
        entangling gates in a cut. MFIB~\cite{markov_quantum_2018} also
        introduced a technique to ``match'' the target fidelity $f$ of the NISQ
        device, which actually reduces the classical computation cost by a
        factor $f$. By matching the fidelity of a realistic quantum hardware ($f
        = 0.51\%$), MFIB was able to simulate $7\times 7$ and $7\times 8$ grids
        with depth $1+40+1$ by numerically computing $10^6$ amplitudes in
        respectively \num{582000} and \num{1407000} core hours on Google Cloud.
        However, MFIB becomes less efficient than qFlex for grids beyond
        $8\times 8$ qubits because of memory requirements. In principle, one
        could mitigate the memory requirements by either using distributed
        memory protocols like \texttt{MPI}, or by partitioning the RQCs in more
        sub-circuits.  Nevertheless, this has been impractical so far. Moreover, the
        low arithmetic intensity intrinsic to this method (which relies on the
        direct evolution of subcircuits) makes it not scalable for
        flop-oriented heterogeneous architectures like Summit.

\end{itemize}

\subsection{qFlex}

In 2018, NASA and Google implemented a circuit simulator -- qFlex -- to compute
amplitudes of arbitrary bitstrings and ran it on NASA’s Electra and Pleiades
supercomputers\cite{villalonga2018flexible}. qFlex novel algorithmic design
emphasizes communication avoiding and minimization of memory footprint, and it
can also be reconfigured to optimize for local memory bandwidth. The
computational resources required by qFlex for simulating RCS are proportional to
the fidelity of the NISQ device \cite{markov_quantum_2018,
villalonga2018flexible}. qFlex provides a fast, flexible and scalable approach
to simulate RCS on CPUs (see Section~\ref{sec:innovation}). In this work, 
qFlex has been redesigned and reimplemented in collaboration with the
Oak Ridge National Laboratory in order to be able to utilize the GPU-accelerated
Summit HPC architecture efficiently (see Section \ref{sec:innovation}). The
computational results presented here, achieving more than 90\% of peak
performance efficiency and 68\% sustained performance efficiency on the World's
most powerful classical computer, define the frontier for quantum computing to
declare Quantum Supremacy. Moreover, given configurability of qFlex as a
function of RAM and desired arithmetic intensity, it can be used to benchmark
not only NISQ computers, but also classical high-end parallel computers.

\section{Innovation Realized}
\label{sec:innovation}

qFlex is a flexible RQC simulator based on an innovative tensor contraction
algorithm for classically simulating quantum circuits that were beyond the reach
of previous approaches \cite{villalonga2018flexible}. More precisely, qFlex is
by design optimized to simulate the generic (worst) case RQCs implemented on
real hardware, which establish the baseline for the applications of near-term
quantum computers.  Moreover, qFlex is agnostic with respect to the randomness
in the choice of single-qubit gates of the RQCs. Therefore, it presents no
fluctuations in performance from one circuit to another in a given ensemble.

For the sake of simplicity, from now on we will focus on planar NISQ devices
where qubits are positioned on a grid, with the only entangling gates being
${\rm CZ}$ gates between two adjacent qubits. As RQCs, we use the prescription
described in Refs.~\cite{boixo_characterizing_2018, new_benchmarks}, which has been
shown to be hard to simulate.

Unlike other approaches, the first step of the qFlex algorithm consists of
contracting the RQC tensor network in the ``time'' direction first. This step
allows us to reduce the RQC to a regular 2D grid of tensors that are then
contracted to produce a single quantum amplitude bitstring (see
Fig.~\ref{fig:cut}). The regularity of the resulting tensors is highly
advantageous for exploiting modern classical computer hardware architectures
characterized by multi-level cache systems, large vector lengths and high level
of parallelism, which requires high arithmetic intensity. The chosen order of
tensor contractions is extremely important for minimization of the dimension of
the largest tensor during the contraction. 

In addition, qFlex expands on a technique introduced in
\cite{chen_64-qubit_2018, li_quantum_2018, chen_classical_2018,
markov_quantum_2018}, which is based on systematic tensor slicing via
fine-grained ``cuts'' that enable us to judiciously balance memory requirements
with the number of concurrent computations. Furthermore, by computing an
appropriate fraction of ``paths'', it is possible to control the ``fidelity'' of
the simulated RCS, and therefore to ``mimic'' the sampling of NISQ devices. Our
simulator can produce a sample of $M$ bitstrings with a target fidelity $f$ at
the same computational cost as to compute $f\cdot M$ noiseless amplitudes.  
More importantly, the use of systematic tensor slicing in our tensor contraction
algorithm results in complete avoidance of communication between MPI processes,
thus making our simulator embarrassingly scalable by design. The number of
tensor ``cuts'' can be properly tuned to fit each independent computation into a
single node. Therefore, all the computations required are run in parallel with
\emph{no communication} between compute nodes. This is an important design
choice that allowed us to avoid the communication bottleneck that would
otherwise highly degrade the performance of qFlex, in particular when GPUs are
used.

Finally, compared to other approaches, qFlex is the most versatile and scalable
simulator of large RCS. In our recent paper \cite{villalonga2018flexible}, qFlex
has been used to benchmark the Google Bristlecone QPU of 72-qubits. To date, our
Bristlecone simulations are the largest numerical computation in terms of
sustained Pflop/s and the number of nodes utilized ever run on NASA Ames
supercomputers, reaching a peak of $20$ Pflop/s (single precision),
corresponding to an efficiency of $64\%$ of the Pleiades and Electra
supercomputers. On Summit, we were able to achieve a sustained performance of
281 Pflop/s (single precision) over the entire supercomputer, which corresponds
to an efficiency of 68\%, as well as a peak performance of 92\%, simulating
circuits of 49 and 121 qubits on a square grid.

To achieve the performance reported here, we have combined several innovations.
Some of them have appeared previously in the literature, as is explained in
detail in their corresponding sections (see Ref.~\cite{villalonga2018flexible}
for more details on the techniques and methods used here).

\subsection{Systematic tensor slicing technique: Communication avoiding and memory footprint reduction}

\begin{figure}
    \centering
    \includegraphics[trim={40pt, 0pt, 40pt, 0pt}, clip, width=45pt]{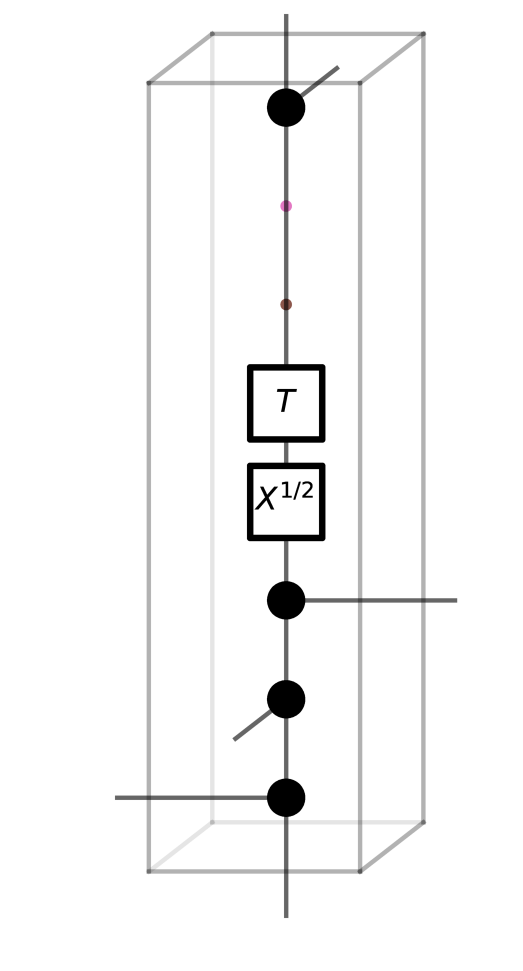}
    \includegraphics[width=170pt]{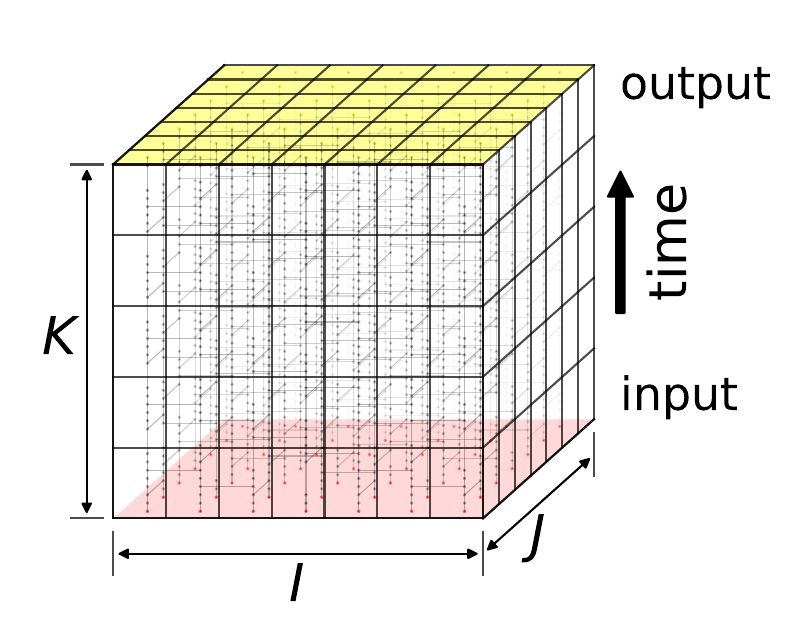}\\
    \includegraphics[width=100pt]{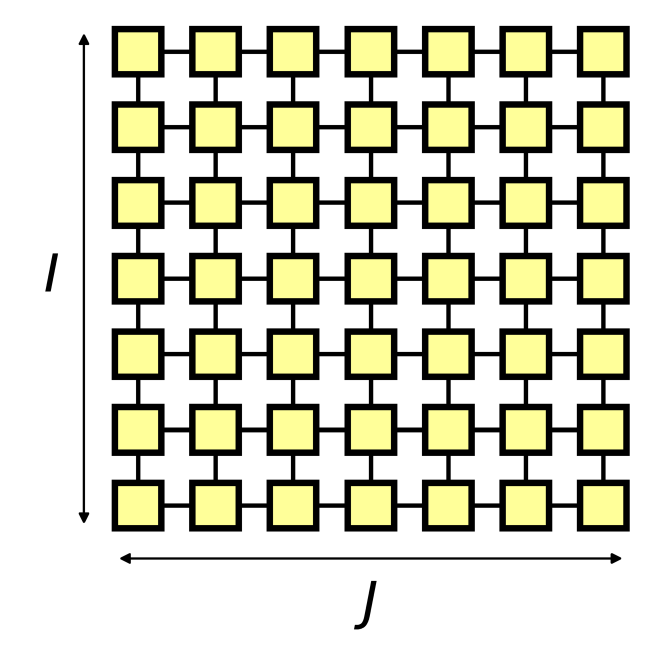}
    \includegraphics[width=105pt]{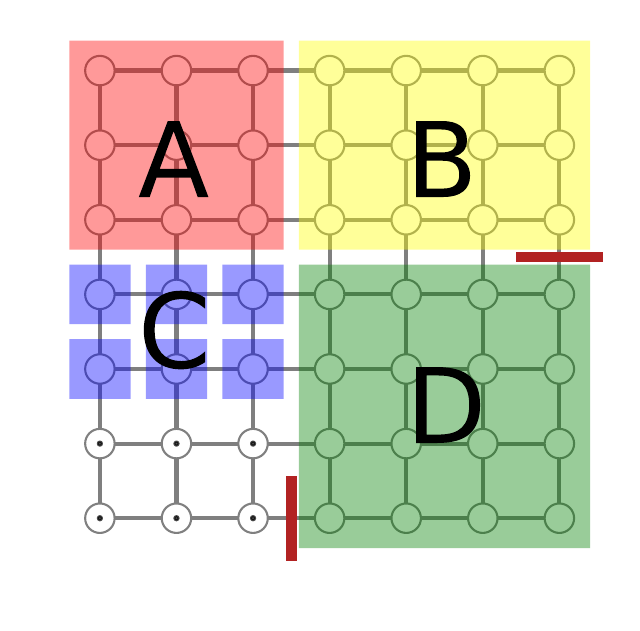}
    \caption{\label{fig:cut}(Top Left): Example of gates (applied in the time
    direction) for a single qubit. (Top Right): RQCs representation as a 3D grid
    of tensors. Each cube is a rank-6 tensor and two adjacent cubes share a
    single bond of dimension 2.  (Bottom Left): 2D network of tensors after
    contracting in ``time'' first. (Bottom Right): Example of cuts applied to
    $7\times7\times(1+40+1)$ RQCs. The tensor network is divided into tensors of
    at most rank 30; once tensors $A$, $B$, and $D$ are contracted with each
    other (with high-arithmetic-intensity) onto tensor $ABD$, $ABD$ is contracted
    with the individual tensors on region $C$, one at a time.. After tensor
    $ABCD$ is formed, the dotted qubits are used for the fast sampling technique
    (see
    Section~\ref{sec:fast}).}
\end{figure}

For RCS with sufficient depth, the contraction of the corresponding 2D grid of
tensors will inevitably lead to the creation of a temporary tensor whose size
exceeds the total amount of memory available in a single node.  For instance,
for a $7\times 7$ RQC of depth $(1+40+1)$, the memory footprint exceeds
the terabytes of memory. To limit the memory footprint
and allow the possibility to contract RQCs on single nodes without
communication, we further expand and optimize the technique of using ``cuts''
\cite{chen_64-qubit_2018, li_quantum_2018, chen_classical_2018, markov_quantum_2018}.

Given a tensor network with $n$ tensors and a set of indexes to contract
$\{i_l\}_{l=1,\ldots}$, $\sum_{i_1,i_2,\ldots} T^1 T^2 \ldots T^n$, we define a
cut over index $i_k$ as the decomposition of the contraction into
$\sum_{i_k}\left(\sum_{\{i_l\}_{l=1,\ldots}-\{i_k\}} T^1 T^2 \ldots T^n\right)$.
This results in the contraction of up to ${\rm dim}(i_k)$ many tensor networks
of lower complexity, namely those covering all slices over index $i_k$ of the
tensors involving that index. The resulting tensor networks can be contracted
independently, which results in a computation that is embarrassingly
parallelizable. It is possible to make more than one cut on a tensor network, in
which case $i_k$ refers to a multi-index, as well as to adaptively choose the
size of the slices. The contribution to the final sum of each of the
contractions (each of the slices over the multi-index cut) is a ``path'', and
the final value of the contraction is the sum of all path contributions.

Fig.~\ref{fig:cut} depicts the cut we used for RQCs of size $7\times 7$ and
depth $(1+40+1)$. The chosen cut allows us to reduce the amount of memory to fit
six contractions on each Summit node (one FPR each GPU). Moreover, as explained
in the next sections, such cuts will greatly improve the sampling of amplitudes
for a given RQC.

\subsection{Fast sampling technique}
\label{sec:fast}

Sampling is a critical feature of random circuit sampling (RCS) and, therefore,
it must be optimized to reduce the time-to-solution.  To this end, we use a
\emph{frugal rejection sampling} which is known to have a low overhead for RCS
\cite{markov_quantum_2018}. In addition, as explained in
Ref.~\cite{villalonga2018flexible}, a faster sampling can be achieved by
recycling most of the large contraction for a given amplitude calculation. More
precisely, if different tensors $A_1(x_1),\,\ldots,\,A_k(x_1)$ and $C(x_2)$ must
be contracted to get the amplitude of bitstring $x = x_1 \bigcup x_2$, our fast
sampling technique first contracts all $A_i(x_1)$ tensors (for a given $x_1$) to
get a tensor $B$. Then, $B$ is recycled and contracted to $C(x_2)$ for many
different $x_2$. Since the most computationally demanding part is the
calculation of $B$, many amplitudes $x = x_1 \bigcup x_2$ can be computed with
different $x_2$ (but the same $x_1$). Since RQCs are extremely chaotic,
arbitrary close bitstrings are uncorrelated \cite{boixo_characterizing_2018}.
These amplitudes are used to perform the frugal rejection sampling of one single
output bitstring in the output sample, and the computation is restarted with a
new random $x_1$ to output the next bitstring. For deep enough RQCs and by
carefully choosing the size of $x_2$, one can reduce the sampling time by an
order of magnitude, as shown in~\cite{villalonga2018flexible}.

\subsection{Noisy simulation}
\label{sec:noisy}

Given the noisy nature of NISQ devices, simulating noisy RCS is of utmost
importance. Intuitively, one would expect that noisier RCS would be easier to
simulate. Indeed, as shown in Ref.~\cite{markov_quantum_2018}, independent
tensor contractions to compute a single amplitude can be seen as orthogonal
``Feynman paths'' (or simply ``paths'').  Since RQCs are very chaotic and
Feynman paths typically contribute equally to the final amplitude, computing
only a fraction $f$ of  paths is equivalent to computing noisy amplitudes of
fidelity $f$. The computational cost for a typical target fidelity of $0.5\%$
for NISQ devices is therefore reduced by a factor of
$200$~\cite{markov_quantum_2018, villalonga2018flexible}.

While computing only a fraction of paths allows for a speedup in the simulation of
noisy devices, computing fewer amplitudes with higher fidelity is an alternative
method to achieve a similar speedup, as we introduced in
Ref.~\cite{villalonga2018flexible}. In particular, computing a fraction $f$
of perfect fidelity amplitudes from a target amplitude set lets us simulate RCS
with fidelity $f$. In general, the fraction of amplitudes computed times their
fidelity equals the fidelity of the simulated sampling.

\subsection{Optimization of the tensor contraction order for optimal time-to-solution}
\label{sec:optimal}

The optimization of the number and position of the cuts and the subsequent
tensor contraction ordering is fundamental for minimizing the total flop count
and the time-to-solution in evaluating a given tensor network and to simulate
sampling.  Note that on modern flop-oriented computer architectures the minimal
flop count does not necessarily guarantee the minimal time-to-solution, making
the optimization problem even more complicated.  Indeed, different tensor
contraction orderings can result in performances that differ by orders of
magnitude. In general, finding the optimal cuts and tensor contraction order is
an NP-Hard problem (closely related to the tree-decomposition problem
\cite{markov_simulating_2008}). However, for planar or quasi-planar tensor
network architectures produced by most of the NISQ devices, such tensor cuts can
be found by carefully splitting the circuits into pieces that minimize the
shared boundary interface; furthermore, we find that choosing a contraction
ordering that prioritizes a few large, high arithmetic intensity contractions
over many smaller, low arithmetic intensity contractions, often provides large
speedups, and therefore better time-to-solution.

\subsection{Out-of-core asynchronous execution of tensor contractions on GPU}
\label{sec:gpupipe}
The novel tensor slicing technique used by the qFlex algorithm removes the
necessity of the inter-process MPI communication and controls the memory
footprint per node, thus paving the way to scalability. In order to achieve high
utilization of the hardware on a heterogeneous Summit node, we implemented an
out-of-core GPU tensor contraction algorithm in the TAL-SH library
\cite{lyakh_tal_sh}, which serves as a computational backend in qFlex. The
TAL-SH library is a numerical tensor algebra library capable of executing basic
tensor algebra operations, most importantly tensor contraction, on multicore CPU
and NVIDIA GPU hardware. The key features of the TAL-SH library that allowed us
to achieve high-performance on GPU-accelerated node architectures are: (a) fast
heterogeneous memory management; (b) fully asynchronous execution of tensor
operations on GPU; (c) fast tensor transpose algorithm; (d) out-of-core
algorithm for executing large tensor contractions on GPU (for tensor
contractions that do not fit into individual GPU memory).

The fast CPU/GPU memory management, and in general fast resource management, is
a necessary prerequisite for achieving high level of asynchronism in executing
computations on CPU and GPU. TAL-SH provides custom memory allocators for Host
and Device memory. These memory allocators use pre-allocated memory buffers
acquired during library initialization. Within such a buffer, each memory
allocator implements a simplified version of the ``buddy'' memory allocator used
in Linux. Since the memory allocation/deallocation occurs inside pre-allocated
memory buffers, it is fast and it is free from highly unwanted side effects
associated with regular CUDA malloc/free, for example serialization of
asynchronous CUDA streams.

All basic tensor operations provided by the TAL-SH library can be executed
asynchronously with respect to the CPU Host on any GPU device available on the
node. The execution of a tensor operation consists of two phases: (a) scheduling
(either successful or unsuccessful, to be re-tried later); (b) checking for
completion, either testing for completion or waiting for completion (each
scheduled tensor operation is a TAL-SH task with its associated TAL-SH task
handle that is used for completion checking). All necessary resources are
acquired during the scheduling phase, thus ensuring an uninterrupted progress of
the tensor operation on a GPU accelerator via a CUDA stream. The tensor
operation scheduling phase includes scheduling the necessary data transfers
between different memory spaces, which are executed asynchronously as well. The
completion checking step also includes consistency control for images of the
same tensor in different memory spaces. All these are automated by TAL-SH.

The tensor contraction operation in TAL-SH is implemented by the general
Transpose-Transpose-GEMM-Transpose (TTGT) algorithm, with an optimized GPU
tensor transpose operation \cite{TensorTranspose1} (see also
Ref.~\cite{TensorTranspose2}). In the TTGT algorithm, the tensor contraction
operation is converted to the matrix-matrix multiplication via permuting tensor
indexes. The performance overhead associated with the tensor transpose steps can
be as low as few percent, or even lower for highly arithmetically intensive
tensor contractions, when using the optimized implementation. Also, all required
Host-to-Device and Device-to-Host data transfers are asynchronous, and they
overlap with kernels execution in other concurrent CUDA streams, thus minimizing
the overhead associated with CPU-GPU data transfers. Additionally, TAL-SH
provides an automated tensor image consistency checking mechanism, enabling
tensor presence on multiple devices with transparent data consistency control
(tensor image is a copy of the tensor on some device).

In this work, we used the regular complex FP32 CGEMM implementation provided by
the cuBLAS library, without tensor core acceleration. The possibility of
reducing the input precision to FP16 in order to use Volta's tensor cores is
strongly conditioned by the precision needed for simulating large quantum
circuits. The average squared output amplitude of a quantum circuit of $n$
qubits is $1/2^n$, which for $n=49$ (simulated in this work) is of the order of
$10^{-15}$ and for $n=121$ (also simulated here) is of the order of $10^{-37}$.
We avoid the possibility of underflowing single precision by normalizing the
input tensors, and renormalizing them back at the end of the computation.

Finally, an important extension to the basic TAL-SH functionality, which was the
key to achieving high performance in this work, is the out-of-core algorithm for
executing large tensor contractions on GPU, that is, for executing tensor
contractions which do not fit into individual GPU memory. Although the Summit
node has IBM AC922 Power9-Volta architecture with unified CPU-GPU coherent
memory efficiently synchronized via NVLink, we did not rely on the unified
memory abstraction provided by the CUDA framework for two reasons. First,
although the hardware assisted memory page migration is very efficient on this
architecture, we believe that the explicitly managed heterogeneous memory and
data transfers deliver higher performance for tensor contractions (this is an
indirect conclusion based on the performance of similar algorithms on Summit).
Second, the TAL-SH library prioritizes performance portability among other
things, that is, it is meant to deliver high performance on other accelerated
HPC systems as well, many of which do not have hardware-assisted coherence
mechanism between CPU and GPU.

The out-of-core tensor contraction algorithm implemented in TAL-SH is based on
recursive decomposition of a tensor operation (tensor contraction in this case)
into smaller tensor operations (tensor contractions) operating on slices of the
original tensors, following the general philosophy presented in
Ref.~\cite{DSVP}. During each decomposition step, the largest tensor dimension
associated with the largest matrix dimension in the TTGT algorithm is split in
half (in TTGT algorithm multiple tensor dimensions are combined into matrix
dimensions, thus matricizing the tensors). This ensures maximization of
arithmetic intensity of individual derived tensor contractions, which is
important for achieving high performance on modern flop-oriented computer
architectures, like the NVIDIA Volta GPU architecture employed in this work
(arithmetic intensity is the flop-to-byte ratio of a given operation). The
recursive decomposition is repeated until all derived tensor operations fit
within available GPU resources. Then the final list of derived tensor operations
is executed by TAL-SH using a pipelined algorithm in which multiple tensor
operations are progressed concurrently, thus overlapping CPU and GPU elementary
execution steps. In general, TAL-SH allows tuning of the number of concurrently
progressed tensor operations, but in this work we restricted it to 2. Each
tensor operation has 5 stages of progress:
\begin{enumerate}
    \item Resource acquisition on the execution device (GPU);
    \item Loading input tensor slices on Host (multithreaded);
    \item Asynchronous execution on GPU: Fast additional resource
          acquisition/release, asynchronous data transfers, asynchronous tensor
          transposes, asynchronous GEMM;
    \item Accumulating the output tensor slice on Host (multithreaded);
    \item Resource release on the execution device (GPU).
\end{enumerate}\vspace{2pt}
The pipelined progressing algorithm is driven by a single CPU thread and is
based on the ``asynchronous yield'' approach, that is, each active (concurrent)
tensor operation proceeds through its consecutive synchronous stages
uninterrupted (unless there is a shortage in available resources, in which case
it is interrupted), but it ``yields'' to the next concurrent tensor operation
once an asynchronous stage starts. The asynchronous GPU execution step, mediated
by the CUDA runtime library, involves asynchronous Host-to-Device and
Device-to-Host data transfers, asynchronous (optimized) tensor transpose kernels
and asynchronous (default) CGEMM cuBLAS calls. Since each NVIDIA Volta GPU has multiple
physical data transfer engines, all incoming and outgoing data transfers are
overlapped with the CUDA kernels execution in different CUDA streams, thus almost
completely removing CPU-GPU data transfer overhead. Overall, this algorithm,
combined with fast tensor transpose and efficient cuBLAS GEMM implementation,
results in highly efficient execution of large tensor contractions on GPU, as
demonstrated by our results (up to 96\% of Volta's theoretical flop/s peak).

\section{How Performance Was Measured}

We used an analytical approach to measuring the number of floating point
operations performed by our algorithm. Our computational workload consists of a
series of tensor contractions operating on dense tensors of rank 1 to 10. Thus,
the total number of Flops executed by an MPI process is the sum of the flop
counts of each individual tensor contraction executed by that MPI process. Each
individual tensor contraction has a well defined flop count:
\begin{equation}
    F = 8 \sqrt{v_0 \cdot v_1 \cdot v_2}
\end{equation}
where $v_0$, $v_1$, $v_2$ are the volumes of the tensors participating in the
tensor contraction (volume is the number of tensor elements in a tensor), and
the factor of 8 shows up due to the use of the complex multiply-add operation (4
real multiplications plus 4 real additions). All tensors used in our simulations
are of complex single precision (complex FP32), SP.

For the timing of the simulations, we do not include job launching,
\texttt{MPI\_Init()}, nor the initialization of the TAL-SH library on each MPI
process, as well as the final time taken to write the results to file by the
master MPI process (scheduler). In order to do so, the two timings are reported after two
\texttt{MPI\_Barrier()} synchronization calls.

Based on the analytical flop count calculation, we have computed multiple
performance metrics in this work. The average SP flop/s count, $C_{A}$, is
computed by dividing the total flop count of our simulation by its execution
time. The peak SP flop/s count, $C_{P}$, is computed by dividing the total flop
count of the largest tensor contraction by the sum of its execution times on
each node, times the number of nodes. Additionally, we have computed the energy
efficiency of the Summit hardware with respect to our simulation. The average
flop/s/watt value, $E_{A}$, is computed by dividing the total flop count of our
simulation by the total energy consumed by Summit during the simulation. This
energy consumption value is actually the upper bound because it includes all
components of Summit, some of which, like disk storage, were not actively used
by our simulation. The Summit HPC system at Oak Ridge National Laboratory has a
total power capacity of 14 MW available to achieve a design specification of
slightly more than 200 Pflop/s peak double-precision performance, and more than
400 Pflop/s single-precision performance. This energy powers 4608 GPU-enabled
compute nodes. Each Summit node contains 2 Power9 CPU, 6 NVIDIA Volta V100 GPUs,
512 GB of DDR4 RAM, 16 GB of HBM2 on-chip GPU memory in each GPU, 1.6 TB of
non-volatile NVMe storage, and 2 physical network interface cards, with a total
power cap of approximately 2400W, of which up to 1800W are consumed by the 6 NVIDIA
Volta GPUs.

\section{Performance Results}

\begin{figure}
    \centering
    \includegraphics[width=210pt]{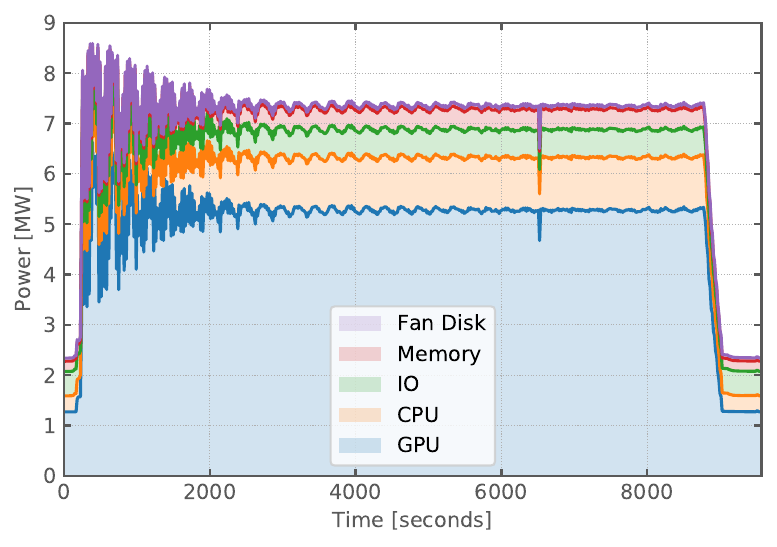}\\
    \includegraphics[width=210pt]{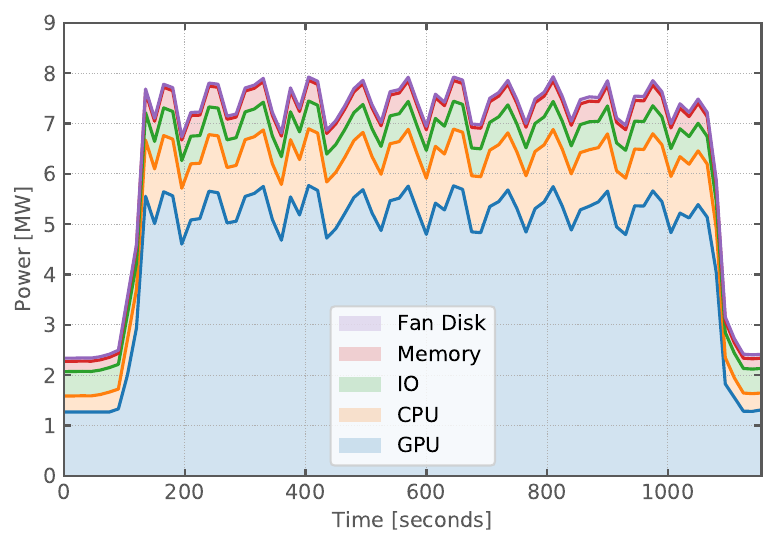}
    \caption{\label{fig:power} Power consumption of Summit, at a given time,
    during our simulations. Note that most of the power is consumed by the GPUs. (Top) $7\times 7\times (1+40+1)$ over 4600 nodes.
    (Bottom) $11\times 11 \times (1+24+1)$ over 4550 nodes.}
\end{figure}

\begin{table*}
\centering
\begin{tabular}{|c|c|c|c|c|c|c|c|c|c|}
\cline{4-5} \cline{6-7}
\multicolumn{1}{c}{} & \multicolumn{1}{c}{} &  & \multicolumn{2}{c|}{\textbf{PFlop/s}} & \multicolumn{2}{c|}{\textbf{Efficiency (\%)}} & \multicolumn{1}{c}{} & \multicolumn{1}{c}{} & \multicolumn{1}{c}{}\tabularnewline
\hline 
\textbf{Circuit Size} & \textbf{Nodes Used} & \textbf{Runtime (h)} & \textbf{Peak} & \textbf{Sust.} & \textbf{Peak} & \textbf{Sust.} & \textbf{Power (MW)} & \textbf{Energy Cost (MWh)} & \textbf{PFlop/s/MW}\tabularnewline
\hline 
\hline 
$7\times7\times(1+40+1)$ & 2300 & 4.84 & 191 & 142 & 92.0 & 68.5 & - & - & -\tabularnewline
$7\times7\times(1+40+1)$ & 4600 & 2.44 & 381 & 281 & 92.1 & 68.0 & 8.075 & 21.1 & 34.8\tabularnewline
$11\times11\times(1+24+1)$ & 4550 & 0.278 & 368 & 261 & 89.8 & 63.7 & 7.3 & 2.32 & 35.8\tabularnewline
\hline 
\end{tabular}
\caption{\label{table:runs}Performance of the simulation of our three runs on
Summit. For the $7\times 7\times (1+40+1)$ circuit we computed $1.01$ million
amplitudes with fidelity $6/1024 \approx 0.5\%$ on each run, while for the
$11\times 11\times (1+24+1)$ circuit we computed 325000 amplitudes of fidelity
$1/64$, which equivalently allows us to sample with fidelity $f\approx 0.5\%$
(see Section~\ref{sec:noisy}). The performance is compared against the
theoretical peak performance, where a maximum of single-precision 15 Tflop/s is
considered for each NVIDIA Volta GPU, and then extrapolated to the number of
nodes used (with 6 GPUs per node). The results show a stable sustained
performance between the simulation of the $7\times 7 \times (1+40+1)$ circuit on
2300 (50\% of Summit) and 4600 nodes (100\% of Summit). The (average) power
consumption reported was measured for the entire Summit, and is therefore not
considered when running on half of the nodes.} 
\end{table*}

We have simulated RCS on two large hard RQCs running on
either half or the entire Summit supercomputer. Both circuits belong to the
class of revised, worst-case circuits on planar graphs (no design weaknesses
exploited)~\cite{markov_quantum_2018} and can be found in~\cite{new_benchmarks}
(\texttt{inst\_7x7\_41\_0.txt} and \texttt{inst\_11x11\_25\_0.txt}). See
Table~\ref{table:runs} for a summary of our performance results.

\subsection{GPU results and peak performance}
\label{sec:gpu_results}

Most of the simulation time is spent in arithmetically intensive, out-of-core
contractions of large tensors (with volumes as large as $v_0=v_1=v_2=2^{30}$)
that do not necessarily fit in the GPU memory. On these contractions, we achieve
a performance efficiency of over $90\%$ with respect to the theoretical
single-precision peak of 15 Tflop/s for the NVIDIA Volta V100 GPUs, and this
efficiency goes as high as 96\%. For this reason, we compute our peak
performance as the average performance of these largest tensor contractions
times the number of GPUs corresponding to the number of nodes used in each
simulation. We therefore achieve a peak performance of \textbf{92\% (381 Pflop/s
when running on 100\% of Summit)} in the $7\times 7\times (1+40+1)$ simulation,
and 90\% (367 Pflop/s when running on 99\% of Summit) in the $11\times 11\times
(1+24+1)$ simulation.

\subsection{Sustained performance}
\label{sec:performance}

While we successfully slice (``cut'') the tensors in our circuits and contract
them in an ordering that leaves most of the computation to large arithmetically
intensive tensor contractions, it is inevitable to deal with a non-negligible
number of less arithmetically intensive tensor contractions, which achieve a
suboptimal performance. Averaged over the entire simulation time, we reach a
sustained performance efficiency of about \textbf{68\% (281 Pflop/s on 100\% of
Summit)} for the $7\times 7\times (1+40+1)$ simulation, and about 64\% (261
Pflop/s on 99\% of Summit) for the $11\times 11\times (1+24+1)$ simulation.
Compared to our previous simulations of random quantum circuits ran on the CPU-only
Pleiades/Electra HPC systems at NASA~\cite{villalonga2018flexible}, the GPU-accelerated
Summit node (2 CPU + 6 GPU) delivers about 45.6 speed-up in running qFlex as compared to
the performance of the averaged Pleiades/Electra dual-CPU node (Pleiades/Electra
HPC systems consist of dual-socket CPU nodes with different families of Intel CPUs
which we average over). The actual single-precision performance delivered by the
averaged Pleiades/Electra node is 1.34 Tflop/s/node, compared to 61.09 Tflop/s/node
sustained by the heterogeneous Summit node. The speed-up is even higher for the peak
performance achieved on both systems.
\subsection{Scaling}
\label{sec:scaling}

Due to the communication-avoiding design of our algorithm, the impact of
communication on scaling is negligible, and therefore the performance is stable
as a function of the number of nodes used, demonstrating an excellent strong
scaling (see Table~\ref{table:runs}).

\subsection{Energy consumption}
\label{sec:energy_consumption}

We report the energy consumed by Summit in our full scale simulations. We
achieve a rate of 34.8 Pflop/s/MW for the $7\times 7\times (1+40+1)$ simulation,
and 35.8 Pflop/s/MW for the $11\times 11\times (1+24+1)$ simulation. Note that
the power consumed should be taken as an upper bound, since it takes into
account the consumption of the entire machine, including components that were
not used in the simulations. Furthermore, we report the energy consumed by the
entire job, including job launch and initialization of the TAL-SH library, which
again slightly lifts the upper bound; this might be part of the reason for
obtaining a slightly larger flop per energy rate on the shorter simulation
($11\times 11\times (1+24+1)$, see Fig.~\ref{fig:power}), where the actual
computing time is smaller in relative terms.

\begin{table*}
\centering
\begin{tabular}{|c|c|c|c|c|c|c|c|}
\cline{3-5} \cline{6-8}
\multicolumn{1}{c}{} &  & \multicolumn{3}{c|}{\textbf{Runtime (hours)}} & \multicolumn{3}{c|}{\textbf{Energy Cost (MWh)}}\tabularnewline
\hline 
\textbf{Circuit Size} & \textbf{Target Fidelity (\%)} & \textbf{Electra} & \textbf{Summit} & \textbf{QPU} & \textbf{Electra} & \textbf{Summit} & \textbf{QPU}\tabularnewline
\hline 
\hline 
$7\times7\times(1+40+1)$ & 0.5 & 59.0 & 2.44 & 0.028 & 96.8 & 21.1 & $4.2\times10^{-4}$\tabularnewline
\hline 
\end{tabular}
\caption{\label{table:classical}Estimated runtimes and energy cost for the sampling 
of $10^6$ amplitudes with fidelity close to $0.5\%$ on NASA
Electra, ORNL Summit, and a superconducting NISQ device (QPU). For the QPU, we
assume a sampling rate of 10 kHz and a power consumption of 15kW (see
Section~\ref{sec:energy_advantages}). Note that the large improvements on
time-to-solution and energy consumption of the QPU as compared to
state-of-the-art classical supercomputers do not extrapolate to other
applications. However, the ``hello world'' nature of the sampling problem
establishes the baseline for the applicability of quantum computers.}
\end{table*}

\section{Implications and Conclusions}

As we explore technologies beyond Moore's law, and in particular as we enter the
NISQ era with quantum devices that surpass classical
capabilities~\cite{arute2019quantum}, it is important to understand the
potential advantages of quantum computing both in terms of time-to-solution and
energy consumption. The research presented here also provides a comparison of
different classical computing architectures, comparing Random Circuit Sampling
implementations on both CPU and GPU based supercomputers (see Table~\ref{table:classical}).

In conclusion, the implications of the proposed research are the following:

\begin{itemize}[label={-}, leftmargin=\parindent]
  \item\textbf{Establishes minimum hardware requirements for quantum computers
        to exceed available classical computation.} qFlex is able to calculate
        the required fidelity, number of qubits, and gate depth required for a
        quantum computer to exceed the computational ability of the most
        powerful available supercomputer for at least one well defined
        computational task: RCS. The demonstration of quantum computational
        supremacy of Ref.~\cite{arute2019quantum} well surpasses this
        threshold.\vspace{4pt}
  \item\textbf{Objectively compares quantum computers in terms of computational
        capability.} Different architectures (e.g. ion traps vs superconducting
        qubits) vary in:
        \begin{enumerate}[leftmargin=20pt]
          \item Number of qubits
          \item Types of gates
          \item Fidelity
          \item Connectivity
          \item Number of qubits per multi-qubit operation
          \item Number of parallel executions
        \end{enumerate}
        RCS allows comparisons across different qubit architectures estimating
        the equivalent amount of classical computation with commonly used
        metrics.\vspace{4pt}
  \item\textbf{Establishes a multi-qubit benchmark for non-Clifford gates.}
        Prior to XEB, there was no benchmarking proposal to measure multiqubit
        fidelity for universal (non-Clifford) quantum gate sets. qFlex enables
        XEB on a large number of qubits and large circuit depths by efficiently
        using classical computing resources. XEB allows quantum hardware vendors
        to calibrate their NISQ devices.\vspace{4pt}
  \item\textbf{Objectively compares classical computers for simulating large
        quantum many-body systems}. Using RCS as a benchmark for classical
        computers, one can compare how different classical computers perform
        simulation of large quantum many-body systems across different classical
        computational architectures. In this paper, we specifically compare
        NASA's Electra supercomputer, which is primarily powered by Intel
        Skylake CPUs, with ORNL's Summit supercomputer, which is primarily
        powered by NVIDIA Volta GPUs.\vspace{4pt}
  \item\textbf{Objectively compares computer energy consumption requirements
        across a variety of quantum and classical architectures for one specific
        computational task.} We estimated the energy cost to produce a sample of
        $10^6$ bistrings for 2 different circuits across classical CPU, GPU, and
        superconducting quantum computer architectures (see
        Table~\ref{table:classical}). The classical computers would take 96.8 MWh
        and 21.1 MWh for NASA's Electra and ORNL's Summit, respectively.  The
        factor of almost 5 of improvement of Summit over Electra is due to
        Summit's GPU architecture.  Comparing this to a superconducting quantum
        computer with a sampling rate of 10 kHz, a quantum computer would have
        an additional 5 orders of magnitude improvement (see
        Table~\ref{table:classical}).  This separation in energy consumption
        performance is much greater than the time-to-solution performance
        improvement of 2 orders of magnitude.  Furthermore, as the number of
        qubits grows we expect the separation in energy performance to continue
        to grow faster than the time-to-solution performance. We emphasize that
        RCS is a computational task particularly favorable to quantum computers
        and an advantage is presently not achievable in practice for most other
        computational problems.\vspace{4pt} 
  \item\textbf{Guides near-term quantum algorithm and hardware design}. The
        simulation capabilities of qFlex can be harnessed for such tasks as
        verifying that a hardware implementation of an algorithm is behaving as
        expected, evaluating relative performance of quantum algorithms that
        could be run on near-term devices, and suggesting hardware architectures
        most suitable for experimenting with specific quantum algorithms in the
        near term. The specific capabilities of qFlex are particularly useful
        for evaluating quantum circuits large enough that the full quantum state
        cannot be produced but for which qFlex can still compute the probability
        of individual outputs of interest. In the case of RCS, it suffices to
        compute the amplitudes of random bistrings to perform rejection sampling
        in order to produce a sample of bistrings \cite{markov_quantum_2018,
        villalonga2018flexible}. In other cases, it might be possible to use
        classical heuristics to determine what probabilities to calculate.
\end{itemize}

\section*{Acknowledgments}

We are grateful for support from NASA Ames Research Center, NASA
Advanced Exploration systems (AES) program, and NASA Transformative Aeronautic Concepts Program (TACP),
and also for support from AFRL Information Directorate under grant
F4HBKC4162G001.
TSH acknowledges support from the Department of Energy Office of Science Early Career Research Program. 
This research used resources of the Oak Ridge Leadership Computing Facility,
which is a DOE Office of Science User Facility supported under Contract
DE-AC05-00OR22725.
We would like to thank Jack Wells, Don Maxwell, and Jim Rogers for their help in
making Summit simulations possible and for providing hardware utilization
statistics.
This manuscript has been authored by UT-Battelle, LLC under Contract No.
DE-AC05-00OR22725 with the U.S. Department of Energy. The United States
Government retains and the publisher, by accepting the article for publication,
acknowledges that the United States Government retains a non-exclusive, paid-up,
irrevocable, world-wide license to publish or reproduce the published form of
this manuscript, or allow others to do so, for United States Government
purposes. The Department of Energy will provide public access to these results
of federally sponsored research in accordance with the DOE Public Access Plan.
(http://energy.gov/downloads/doe-public-access-plan).

\appendix
\subsection{Comparison with variable elimination algorithms}
\label{sec:variable_elimination}

As discussed in Section~\ref{sec:soa}, variable elimination algorithms over
undirected graphical models can be used to simulate quantum circuits, which has
been successfully exploited in Refs.~\cite{boixo_simulation_2017,
chen_classical_2018, zhang2019alibaba}. In this appendix we estimate the
computational resources needed by such algorithms to simulate the circuits
presented in this work. In our estimates, we make use of
the algorithm and benchmarks presented in Refs.~\cite{chen_classical_2018,
zhang2019alibaba}.\\

\subsubsection{Algorithm}

The main idea is to represent the tensor network (TN) of a quantum circuit as a
graph closely related to its line graph: an undirected graphical model. Each
vertex in the graph corresponds to a unique index in the TN, and is therefore a
binary variable. For each tensor or gate in the TN shared by a set of vertices,
edges are added to the graph so that such vertices are connected all-to-all,
forming a clique (complete subgraph). By representing TNs this way, one can take
advantage of the structure of particular gates and explicitly simplify the
problem by obtaining a simpler graph. For example, diagonal two-qubit gates have
only two distinct variables (vertices in the graph), unlike general two-qubit
gates, which have four; in addition, diagonal one-qubit gates have a single
variable, instead of two. This representation is equivalent to TNs if these are
allowed to have hyperedges, corresponding to the situation where a vertex
belongs to more than two cliques (tensors) simultaneously.

Once the undirected graphical model corresponding to a circuit is built, the
contraction is carried out in a similar way as a TN contraction, which is
referred to as a variable elimination procedure. We will use the terms
contraction and elimination interchangeably. The variable elimination over the
entire graph proceeds as follows. First, an elimination ordering is chosen. Each
variable being eliminated corresponds to the contraction of an index from a TN
perspective, and results in the elimination of the variable from the graph, as
well as the subsequent formation of a clique between all cliques which contained
the variable prior to its elimination. This way, as the elimination procedure
continues, cliques of different sizes are formed; we define the
\emph{contraction width} of the ordering as the size of the largest clique
formed along the process. The size of the largest tensor formed along the
contraction is equal to $2^{\text{contraction width}}$. At the same time, the
time complexity of the elimination of a variable is proportional to $2^{c}$,
where $c$ is the size of the resulting clique, and so the time complexity of the
entire elimination process is proportional to $\sum_{i} 2^{c_i}$, where $i$
labels all variables, and is dominated by the term $2^{\text{contraction
width}}$. Given an undirected graphical model, finding the elimination ordering
that minimizes the contraction width is an NP-hard problem; such optimal
contraction width is called the \emph{treewidth} (TW) of the graph. There are
heuristic algorithms to find suboptimal elimination orderings, as well as
computationally intensive algorithms that guarantee optimality, which also
provide suboptimal solutions when run for a finite amount of time.
Ref.~\cite{boixo_simulation_2017} introduced the use of {\tt QuickBB} for
finding good elimination orderings for the simulation of quantum circuits.

If the TW of the graph (or its contraction width found, for practical purposes)
is so large that the amount of memory needed for the simulation exceeds
computational resources, it is still possible to perform the simulation, as
introduced in Ref.~\cite{chen_classical_2018}. The idea here is that by
projecting a single variable of the graph to either one of its two values, the
resulting subgraph (which has lost the vertex corresponding to the variable, as
well as all edges adjacent to it) might require substantially lower
computational resources, at the expense of doubling the number of graphs we have
to evaluate or contract. By choosing a careful subset of variables to project,
the resulting subgraphs are easy enough to contract, and their evaluations are
independent of each other, making the simulation embarrassingly parallelizable.
The projection of a number $p$ of variables leads to the evaluation of $2^p$
circuits, all of which have decreased their time complexity by some factor
common to all, and have possibly also decreased their contraction width.
Ref.~\cite{chen_classical_2018} provides a heuristic algorithm to choose this
subset of $p$ variables that works well in practice. First, {\tt QuickBB} is run
over the initial graph for $S$ seconds in order to find a good elimination
ordering. Then, the contraction costs of the subgraphs that would result from
the projection of each variable are computed, and the variable that decreases
the resulting cost is chosen to be projected. In doing this, the elimination
ordering inherited from the previous step is used. The same procedure is
greedily carried out $p$ times. Finally, {\tt QuickBB} is run over the final
graph (where $p$ variables have been projected) in an attempt to find a
potentially better elimination ordering than the one inherited from the first
step; this might in some cases not decrease the elimination cost.\\

\subsubsection{Methodology}

\begin{figure}
    \centering
    \includegraphics[width=210pt]{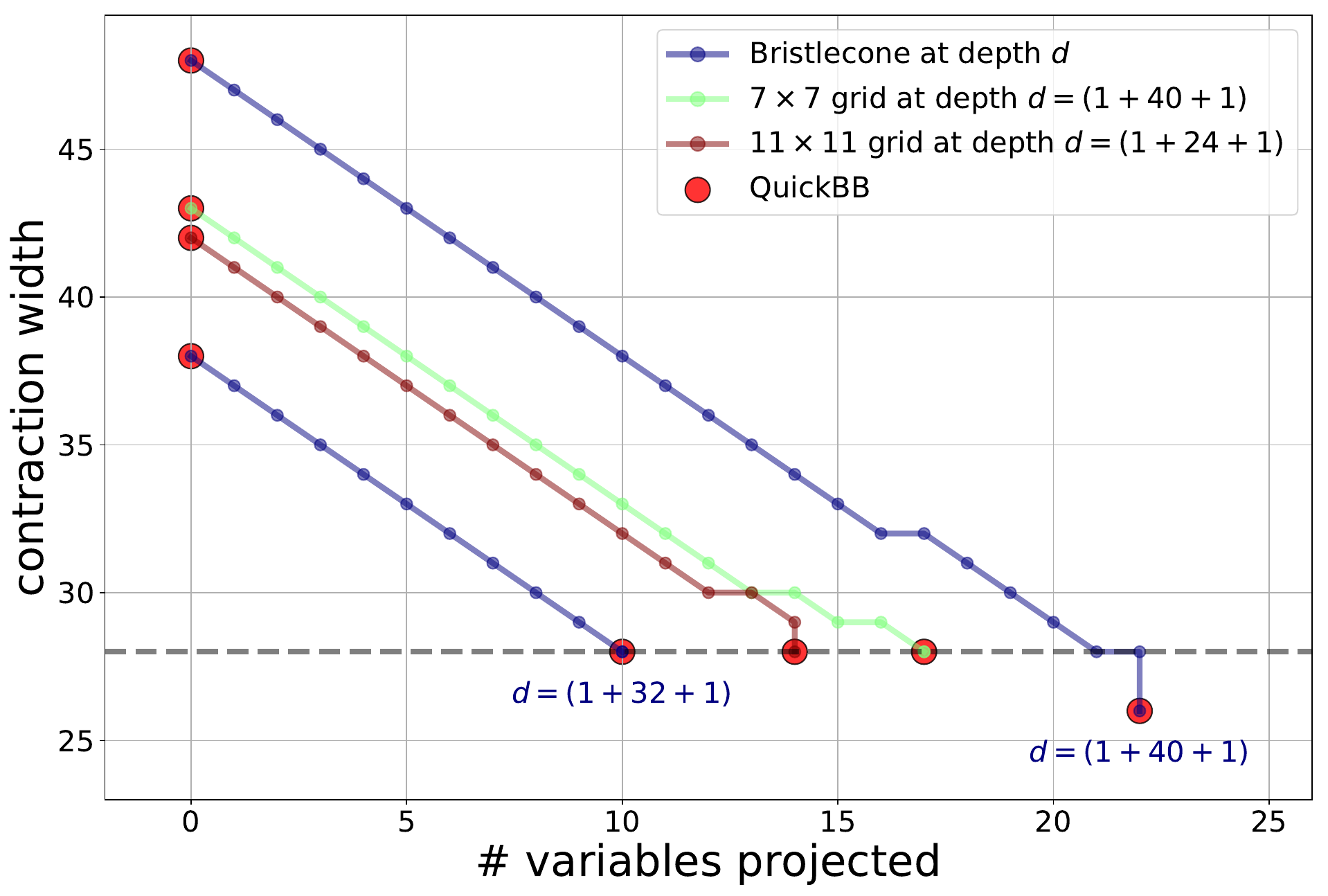}\\
    \caption{\label{fig:projections} Contraction widths as a function of the
        number of variables projected following the algorithm of
        Refs.~\cite{chen_classical_2018, zhang2019alibaba}. Both circuits
        simulated in this work have a complexity that lays in between
        Bristlecone at depths $(1+32+1)$ and $(1+40+1)$, which were benchmarked
        in Ref.~\cite{zhang2019alibaba} by computing 1K amplitudes and 1
        amplitude, respectively.}
\end{figure}

\begin{table*}
    \centering
    \begin{tabular}{|c|c|c|c|c|c|}
    \hline 
    \textbf{Simulator} & \textbf{Circuit instance} &
    \textbf{Num. variables projected, $p$} & \textbf{Cost, $\sum_i 2^{c_i}$} & \textbf{Runtime, $T$ (h)} & \textbf{Memory requirements (PB)}\tabularnewline
    \hline 
    \hline 
    Alibaba & Bristlecone$\times(1+32+1)$ & ${10}$ & $2.82 \times 10^{9}$ & 0.0293 & 4.72 \tabularnewline
    Alibaba & Bristlecone$\times(1+40+1)$ & ${22}$ & $9.11 \times 10^{8}$ & 38.8 & 4.72 \tabularnewline
    Alibaba & $7\times7\times(1+40+1)$ & ${17}$ & $2.48 \times 10^{9}$ & 3.31 & 4.72 \tabularnewline
    Alibaba & $11\times11\times(1+24+1)$ & ${14}$ & $5.04 \times 10^{9}$ & 0.838 & 4.72 \tabularnewline
    qFlex & $7\times7\times(1+40+1)$ & - & - & 2.44 & 2.36 \tabularnewline
    qFlex & $11\times11\times(1+24+1)$ & - & - & 0.278 & 2.36 \tabularnewline
    \hline 
    \end{tabular}
    \caption{\label{table:variable_elimination} Comparison of qFlex with the
        variable elimination simulator of Refs.~\cite{chen_classical_2018,
        zhang2019alibaba} (Alibaba). Runtimes for the sampling of 1M bitstrings
        at fidelity $0.5\%$ are considered. We extrapolate the runtimes reported
        in~\cite{zhang2019alibaba} to the circuits benchmarked here, as well as
        to the peak performance of Summit. This implies assuming that the simulator
        of~\cite{zhang2019alibaba} is implementable on GPUs with similar
        efficiency as that one achieved on CPUs. Finally, the fact that fast sampling (see
        Section~\ref{sec:fast}) is not implemented for variable elimination
        algorithms is neglected; considering this fact would increase time
        estimates for this simulator by a factor of between 3 and 10 (see
        Refs.~\cite{markov_quantum_2018, villalonga2018flexible,
        chen_quantum_2019}).} 
\end{table*}

In Ref.~\cite{chen_classical_2018}, this algorithm was used in order to
benchmark the simulation of circuits on square grids of qubits. As explained in
Ref.~\cite{villalonga2018flexible} and in Section~\ref{sec:soa}, the circuits
benchmarked in Ref.~\cite{chen_classical_2018} had gate sequences that made them
easier to simulate than those of Ref.~\cite{villalonga2018flexible} and of the
present work. In Ref.~\cite{zhang2019alibaba}, the authors used a newer version
of their algorithm to simulate circuits on the Bristlecone grid introduced
in~\cite{villalonga2018flexible}, with substantial speedups. In this appendix,
we use these benchmarks in order to estimate the performance of this variable
elimination algorithm in the simulation of the circuits benchmarked here, which
are defined over rectangular grids. For this estimate, we have implemented the
heuristic algorithm of~\cite{chen_classical_2018} for choosing the subset of
variables that are projected. We have run our implementation of this algorithm
on the Bristlecone circuits of Ref.~\cite{zhang2019alibaba} with a value of
$S=1800$ seconds (30 minutes), which we believe decreases the TW of the
resulting graphs enough to fit within the memory resources used, as reported by
the authors. Then, with the same setting of $S=1800$ seconds, we run the
algorithm over the $7\times 7\times (1+40+1)$ and the $11\times 11\times
(1+24+1)$ circuits presented here. After projecting enough variables so that the
memory requirements of the contractions are as low as those found for the
Bristlecone circuits (see Fig.~\ref{fig:projections} for the resulting
contraction widths), we use the following expression to determine the
computation time for the computation of an amplitude:
\begin{align}
\label{eq:elimination_time}
    T = D^{-1} \cdot 2^p \cdot (\text{elim. cost after }p\text{ projections}) / n_{\text{cores}}\text{,}
\end{align}
where $D$ is a constant factor, $p$ is the number of variables projected, and
$n_{\text{cores}}$ is the number of cores used in the computation. Constant $D$
is obtained from the benchmarking data reported in Ref.~\cite{zhang2019alibaba}:
the computation of an amplitude of Bristlecone circuits at depths $(1+32+1)$ and
$(1+40+1)$ took respectively 0.43 seconds and 580.7 seconds when using
$\num{127512}$ CPU cores. In the first case, 10 variables were projected, while
22 were projected in the second. This leads to a value of $D=52.7$ MHz and 51.6
MHz, respectively; we choose to use $D=52.7$ MHz in our estimates since it leads
to slightly lower computation times. Note that the fact that both values of $D$
are extremely close to each other gives us confidence that the $S$ value of {\tt
QuickBB} yields indeed results close to those of Ref.~\cite{zhang2019alibaba}.

In order to have meaningful comparisons between both simulators, we assume that
a number of CPUs similar to Summit in the number of delivered FLOPs is
accessible. For that, note that each of the Intel Xeon 8163 CPUs used in
Ref.~\cite{zhang2019alibaba} has two AVX-512 units that deliver together 160
Gflops @ 2.5 GHz (single precision). Each node has 88 CPU cores (14.08 Tflops)
and in total 1449 nodes are used ($\num{127512}$ cores), for a total theoretical
peak of 20.4 Pflops delivered. Each Summit node delivers a peak of 90 Tflops
(see Section~\ref{sec:gpu_results}), and so it corresponds to 6.4 of the CPU
nodes considered here. All 1449 nodes are performance-wise equivalent to about
226 Summit nodes, although with about twice the amount of memory per flop: 160
GB per CPU node compared to 512 GB per Summit node. We therefore consider that
4600 Summit nodes (100\% of Summit) correspond to $n_{\text{cores}} =
\num{127512} \times 4600 / 226 = \num{2595377}$ CPU cores. Such a number of
cores would require $1449 \times 160 \times 4600 / 226 = 4.72$ PB of memory.\\

\subsubsection{Results}
\label{sec:app_results}

Table~\ref{table:variable_elimination} shows the estimated runtimes and memory
requirements for the simulation of the circuit presented in the current work
using the variable elimination simulator of Ref.~\cite{zhang2019alibaba}. For
comparison, we also include the results of qFlex. For the $7\times 7\times
(1+40+1)$ circuit, qFlex is somewhat faster, and requires about half the memory
resources. For the $11\times 11\times (1+32+1)$, qFlex is about $3\times$
faster, halving also the memory requirements.\\

\subsubsection{Optimistic assumptions on the variable elimination algorithm}

In Section~\ref{sec:app_results}, we have computed time estimates for the
simulation of random circuit sampling of 1M bitstrings at fidelity $0.5\%$.
However, there are two optimistic assumptions on the performance of the variable
elimination simulator that we discuss here. First, sampling 1M bitstrings from
an RQC with fidelity $0.5\%$ involves the computation of 5000 perfect fidelity
\emph{batches} of amplitudes (see Fig.~\ref{sec:innovation}). One of the
disadvantages of variable elimination algorithm of
Refs.~\cite{chen_classical_2018, zhang2019alibaba} compared to the tensor
network techniques described here is its inability to compute batches of
amplitudes of correlated bitstrings at the same cost as a single amplitude
(although it might be possible to implement this technique in the future).
Accounting for this fact should increase the computation time to sample a
bitstring by a multiplicative factor (between 3 and 10, in practice, as shown in
Refs.~\cite{markov_quantum_2018, villalonga2018flexible, chen_quantum_2019}).
Second, we assume that a variable elimination algorithm is successfully
implemented to run efficiently on GPUs, which is in principle hindered by the
low arithmetic intensity of the contraction of variables one at a time; given
such an assumption, runtimes for this algorithm on Summit can be estimated. We
have made these assumptions in order to give what we consider would be the best
possible scenario in which variable elimination algorithms could operate;
estimates should be larger by some factors.

\bibliographystyle{ieeetr}
\bibliography{refs}

\begin{thebibliography}{10}

\bibitem{feynman1982simulating}
R.~P. Feynman, ``{Simulating Physics with Computers},'' {\em International
  Journal Of Theoretical Physics}, vol.~21, no.~6-7, pp.~467--488, 1982.

\bibitem{feynman1985quantum}
R.~P. Feynman, ``Quantum mechanical computers,'' {\em Optics news}, vol.~11,
  no.~2, pp.~11--20, 1985.

\bibitem{barends_superconducting_2014}
R.~Barends, J.~Kelly, A.~Megrant, A.~Veitia, D.~Sank, E.~Jeffrey, T.~C. White,
  J.~Mutus, A.~G. Fowler, B.~Campbell, and {others}, ``Superconducting quantum
  circuits at the surface code threshold for fault tolerance,'' {\em Nature},
  vol.~508, no.~7497, pp.~500--503, 2014.

\bibitem{kelly_state_2015}
J.~Kelly, R.~Barends, A.~G. Fowler, A.~Megrant, E.~Jeffrey, T.~C. White,
  D.~Sank, J.~Y. Mutus, B.~Campbell, Y.~Chen, and {others}, ``State
  preservation by repetitive error detection in a superconducting quantum
  circuit,'' {\em Nature}, vol.~519, no.~7541, pp.~66--69, 2015.

\bibitem{wang201816}
Y.~Wang, Y.~Li, Z.-q. Yin, and B.~Zeng, ``16-qubit ibm universal quantum
  computer can be fully entangled,'' {\em npj Quantum Information}, vol.~4,
  no.~1, p.~46, 2018.

\bibitem{barends2016digitized}
R.~Barends, A.~Shabani, L.~Lamata, J.~Kelly, A.~Mezzacapo, U.~Las~Heras,
  R.~Babbush, A.~G. Fowler, B.~Campbell, Y.~Chen, {\em et~al.}, ``Digitized
  adiabatic quantum computing with a superconducting circuit,'' {\em Nature},
  vol.~534, no.~7606, pp.~222--226, 2016.

\bibitem{arute2019quantum}
F.~Arute, K.~Arya, R.~Babbush, D.~Bacon, J.~C. Bardin, R.~Barends, R.~Biswas,
  S.~Boixo, F.~G. Brandao, D.~A. Buell, {\em et~al.}, ``Quantum supremacy using
  a programmable superconducting processor,'' {\em Nature}, vol.~574, no.~7779,
  pp.~505--510, 2019.

\bibitem{NCbook}
M.~A. Nielsen and I.~L. Chuang, {\em Quantum computation and quantum
  information}.
\newblock Cambridge University Press, Cambridge, 2000.

\bibitem{RPbook}
E.~G. Rieffel and W.~Polak, {\em {Quantum Computing: A Gentle Introduction}}.
\newblock Cambridge, MA: MIT Press, 2011.

\bibitem{grover1996fast}
L.~K. Grover, ``{A fast quantum mechanical algorithm for database search},'' in
  {\em Proceedings of the twenty-eighth annual ACM symposium on Theory of
  computing - STOC '96}, pp.~212--219, ACM, 1996.

\bibitem{aspuru-guzik_simulated_2005}
A.~Aspuru-Guzik, A.~D. Dutoi, P.~J. Love, and M.~Head-Gordon, ``{Chemistry:
  Simulated quantum computation of molecular energies},'' {\em Science},
  vol.~309, pp.~1704--1707, sep 2005.

\bibitem{babbush_low_2017}
R.~Babbush, N.~Wiebe, J.~McClean, J.~McClain, H.~Neven, and G.~K.~L. Chan,
  ``{Low-Depth Quantum Simulation of Materials},'' {\em Physical Review X},
  vol.~8, may 2018.

\bibitem{jiang_quantum_2018}
Z.~Jiang, K.~J. Sung, K.~Kechedzhi, V.~N. Smelyanskiy, and S.~Boixo, ``{Quantum
  Algorithms to Simulate Many-Body Physics of Correlated Fermions},'' {\em
  Physical Review Applied}, vol.~9, p.~44036, apr 2018.

\bibitem{babbush_encoding_2018}
R.~Babbush, C.~Gidney, D.~W. Berry, N.~Wiebe, J.~McClean, A.~Paler, A.~Fowler,
  and H.~Neven, ``{Encoding Electronic Spectra in Quantum Circuits with Linear
  T Complexity},'' {\em Physical Review X}, vol.~8, p.~41015, oct 2018.

\bibitem{shor1994algorithms}
P.~Shor, ``{Algorithms for quantum computation: discrete logarithms and
  factoring},'' in {\em Proceedings 35th Annual Symposium on Foundations of
  Computer Science}, pp.~124--134, Ieee, 1994.

\bibitem{smelyanskiy2018non}
V.~N. Smelyanskiy, K.~Kechedzhi, S.~Boixo, S.~V. Isakov, H.~Neven, and
  B.~Altshuler, ``Non-ergodic delocalized states for efficient population
  transfer within a narrow band of the energy landscape,'' {\em
  arXiv:1802.09542}, 2018.

\bibitem{fowler2012surface}
A.~G. Fowler, M.~Mariantoni, J.~M. Martinis, and A.~N. Cleland, ``Surface
  codes: Towards practical large-scale quantum computation,'' {\em Phys. Rev.
  A}, vol.~86, no.~3, p.~032324, 2012.

\bibitem{boixo_characterizing_2018}
S.~Boixo, S.~V. Isakov, V.~N. Smelyanskiy, R.~Babbush, N.~Ding, Z.~Jiang, M.~J.
  Bremner, J.~M. Martinis, and H.~Neven, ``{Characterizing quantum supremacy in
  near-term devices},'' {\em Nature Physics}, vol.~14, pp.~1--6, jun 2018.

\bibitem{bremner_achieving_2017}
M.~J. Bremner, A.~Montanaro, and D.~J. Shepherd, ``{Achieving quantum supremacy
  with sparse and noisy commuting quantum computations},'' {\em Quantum},
  vol.~1, p.~8, apr 2016.

\bibitem{neill_blueprint_2018}
C.~Neill, P.~Roushan, K.~Kechedzhi, S.~Boixo, S.~V. Isakov, V.~Smelyanskiy,
  A.~Megrant, B.~Chiaro, A.~Dunsworth, K.~Arya, {\em et~al.}, ``A blueprint for
  demonstrating quantum supremacy with superconducting qubits,'' {\em Science},
  vol.~360, no.~6385, pp.~195--199, 2018.

\bibitem{de_raedt_massively_2007}
K.~{De Raedt}, K.~Michielsen, H.~{De Raedt}, B.~Trieu, G.~Arnold, M.~Richter,
  T.~Lippert, H.~Watanabe, and N.~Ito, ``{Massively parallel quantum computer
  simulator},'' {\em Computer Physics Communications}, vol.~176, pp.~121--136,
  jan 2007.

\bibitem{preskill_quantum_2018}
J.~Preskill, ``Quantum computing in the nisq era and beyond,'' {\em Quantum},
  vol.~2, p.~79, 2018.

\bibitem{smelyanskiy_qhipster:_2016}
M.~Smelyanskiy, N.~P.~D. Sawaya, and A.~Aspuru-Guzik, ``{qHiPSTER: The Quantum
  High Performance Software Testing Environment},'' {\em arXiv:1601.07195},
  2016.

\bibitem{haner20170}
T.~H{\"{a}}ner and D.~S. Steiger, ``{0.5 Petabyte Simulation of a 45-Qubit
  Quantum Circuit},'' in {\em Proceedings of the International Conference for
  High Performance Computing, Networking, Storage and Analysis}, p.~33, ACM,
  2017.

\bibitem{aaronson2016complexity}
S.~Aaronson and L.~Chen, ``Complexity-theoretic foundations of quantum
  supremacy experiments,'' {\em arXiv:1612.05903}, 2016.

\bibitem{bouland_quantum_2018}
A.~Bouland, B.~Fefferman, C.~Nirkhe, and U.~Vazirani, ``{Quantum Supremacy and
  the Complexity of Random Circuit Sampling},'' {\em arXiv:1803.04402}, 2018.

\bibitem{pednault_breaking_2017}
E.~Pednault, J.~A. Gunnels, G.~Nannicini, L.~Horesh, T.~Magerlein,
  E.~Solomonik, and R.~Wisnieff, ``{Breaking the 49-Qubit Barrier in the
  Simulation of Quantum Circuits},'' {\em arXiv:1710.05867}, 2017.

\bibitem{chen_64-qubit_2018}
Z.~Y. Chen, Q.~Zhou, C.~Xue, X.~Yang, G.~C. Guo, and G.~P. Guo, ``{64-Qubit
  Quantum Circuit Simulation},'' {\em Science Bulletin}, vol.~63, pp.~964--971,
  aug 2018.

\bibitem{li_quantum_2018}
R.~Li, B.~Wu, M.~Ying, X.~Sun, and G.~Yang, ``{Quantum Supremacy Circuit
  Simulation on Sunway TaihuLight},'' {\em arXiv:1804.04797 [quant-ph]}, apr
  2018.

\bibitem{chen_classical_2018}
J.~Chen, F.~Zhang, C.~Huang, M.~Newman, and Y.~Shi, ``{Classical Simulation of
  Intermediate-Size Quantum Circuits},'' {\em arXiv:1805.01450 [quant-ph]}, may
  2018.

\bibitem{markov_quantum_2018}
I.~L. Markov, A.~Fatima, S.~V. Isakov, and S.~Boixo, ``{Quantum Supremacy Is
  Both Closer and Farther than It Appears},'' {\em arXiv:1807.10749
  [quant-ph]}, jul 2018.

\bibitem{villalonga2018flexible}
B.~Villalonga, S.~Boixo, B.~Nelson, C.~Henze, E.~Rieffel, R.~Biswas, and
  S.~Mandr{\`a}, ``A flexible high-performance simulator for verifying and
  benchmarking quantum circuits implemented on real hardware,'' {\em NPJ
  Quantum Information}, vol.~5, pp.~1--16, 2019.

\bibitem{de_raedt_massively_2018}
H.~{De Raedt}, F.~Jin, D.~Willsch, M.~Nocon, N.~Yoshioka, N.~Ito, S.~Yuan, and
  K.~Michielsen, ``{Massively parallel quantum computer simulator, eleven years
  later},'' {\em arXiv:1805.04708}, 2018.

\bibitem{ibm_2019}
``{Quantum} computation center opens, {IBM} {Research} {Blog},'' September
  2019.

\bibitem{aaronson2011computational}
S.~Aaronson and A.~Arkhipov, ``{The Computational Complexity of Linear
  Optics},'' in {\em Proceedings of the forty-third annual ACM symposium on
  Theory of computing}, pp.~333--342, ACM, 2010.

\bibitem{bremner_average-case_2016}
M.~J. Bremner, A.~Montanaro, and D.~J. Shepherd, ``{Average-Case Complexity
  Versus Approximate Simulation of Commuting Quantum Computations},'' {\em
  Physical Review Letters}, vol.~117, p.~80501, aug 2016.

\bibitem{aaronson2017complexity}
S.~Aaronson and L.~Chen, ``{Complexity-Theoretic Foundations of Quantum
  Supremacy Experiments},'' in {\em LIPIcs-Leibniz International Proceedings in
  Informatics}, vol.~79, Schloss Dagstuhl-Leibniz-Zentrum fuer Informatik,
  2016.

\bibitem{harrow2018approximate}
A.~Harrow and S.~Mehraban, ``Approximate unitary $ t $-designs by short random
  quantum circuits using nearest-neighbor and long-range gates,'' {\em arXiv
  preprint arXiv:1809.06957}, 2018.

\bibitem{movassagh_efficient_2018}
R.~Movassagh, ``{Efficient unitary paths and quantum computational supremacy: A
  proof of average-case hardness of Random Circuit Sampling},'' {\em
  arXiv:1810.04681}, 2018.

\bibitem{bishop2017quantum}
L.~S. Bishop, S.~Bravyi, A.~Cross, J.~M. Gambetta, and J.~Smolin, ``Quantum
  volume,'' {\em Quantum Volume. Technical Report}, 2017.

\bibitem{magesan_robust_2011}
E.~Magesan, J.~M. Gambetta, and J.~Emerson, ``Scalable and robust randomized
  benchmarking of quantum processes,'' {\em Phys. Rev. Lett.}, vol.~106,
  no.~18, p.~180504, 2011.

\bibitem{knill2008randomized}
E.~Knill, D.~Leibfried, R.~Reichle, J.~Britton, R.~Blakestad, J.~Jost,
  C.~Langer, R.~Ozeri, S.~Seidelin, and D.~Wineland, ``Randomized benchmarking
  of quantum gates,'' {\em Phys. Rev. A}, vol.~77, no.~1, p.~012307, 2008.

\bibitem{magesan_characterizing_2012}
E.~Magesan, J.~M. Gambetta, and J.~Emerson, ``Characterizing {Quantum} {Gates}
  via {Randomized} {Benchmarking},'' {\em Phys. Rev. A}, vol.~85, April 2012.

\bibitem{erhard2019characterizing}
A.~Erhard, J.~J. Wallman, L.~Postler, M.~Meth, R.~Stricker, E.~A. Martinez,
  P.~Schindler, T.~Monz, J.~Emerson, and R.~Blatt, ``Characterizing large-scale
  quantum computers via cycle benchmarking,'' {\em arXiv:1902.08543}, 2019.

\bibitem{Britt2017}
K.~A. {Britt}, F.~A. {Mohiyaddin}, and T.~S. {Humble}, ``Quantum accelerators
  for high-performance computing systems,'' in {\em 2017 IEEE International
  Conference on Rebooting Computing (ICRC)}, pp.~1--7, Nov 2017.

\bibitem{Humble2018}
T.~S. Humble, R.~J. Sadlier, and K.~A. Britt, ``Simulated execution of hybrid
  quantum computing systems,'' in {\em Quantum Information Science, Sensing,
  and Computation X}, vol.~10660, p.~1066002, International Society for Optics
  and Photonics, 2018.

\bibitem{gottesman1998heisenberg}
D.~Gottesman, ``The heisenberg representation of quantum computers,'' {\em
  arXiv preprint quant-ph/9807006}, 1998.

\bibitem{aaronson2004improved}
S.~Aaronson and D.~Gottesman, ``Improved simulation of stabilizer circuits,''
  {\em Physical Review A}, vol.~70, no.~5, p.~052328, 2004.

\bibitem{bravyi_improved_2016}
S.~Bravyi and D.~Gosset, ``{Improved Classical Simulation of Quantum Circuits
  Dominated by Clifford Gates},'' {\em Physical Review Letters}, vol.~116,
  p.~250501, jun 2016.

\bibitem{bennink2017}
R.~S. Bennink, E.~M. Ferragut, T.~S. Humble, J.~A. Laska, J.~J. Nutaro, M.~G.
  Pleszkoch, and R.~C. Pooser, ``Unbiased simulation of near-clifford quantum
  circuits,'' {\em Physical Review A}, vol.~95, no.~6, p.~062337, 2017.

\bibitem{markov_simulating_2008}
I.~L. Markov and Y.~Shi, ``{Simulating quantum computation by contracting
  tensor networks},'' {\em SIAM Journal on Computing}, vol.~38, pp.~963--981,
  jan 2008.

\bibitem{biamonte2017tensor}
J.~Biamonte and V.~Bergholm, ``Tensor networks in a nutshell,'' {\em arXiv
  preprint arXiv:1708.00006}, 2017.

\bibitem{TNQVM}
A.~McCaskey, E.~Dumitrescu, M.~Chen, D.~Lyakh, and T.~Humble, ``Validating
  quantum-classical programming models with tensor network simulations,'' {\em
  PLOS ONE}, vol.~13, pp.~1--19, 12 2018.

\bibitem{boixo_simulation_2017}
S.~Boixo, S.~V. Isakov, V.~N. Smelyanskiy, and H.~Neven, ``{Simulation of
  low-depth quantum circuits as complex undirected graphical models},'' {\em
  arXiv:1712.05384 [quant-ph]}, dec 2017.

\bibitem{new_benchmarks}
S.~Boixo and C.~Neill, ``The question of quantum supremacy,'' {\em Google AI
  Blog}, May 2018.
\newblock Available on GitHub at \url{https://github.com/sboixo/GRCS}.

\bibitem{chen_quantum_2019}
M.-C. Chen, R.~Li, L.~Gan, X.~Zhu, G.~Yang, C.-Y. Lu, and J.-W. Pan, ``Quantum
  {Teleportation}-{Inspired} {Algorithm} for {Sampling} {Large} {Random}
  {Quantum} {Circuits},'' {\em arXiv:1901.05003 [quant-ph]}, January 2019.

\bibitem{lyakh_tal_sh}
D.~Lyakh, ``Tensor algebra library routines for shared memory systems
  (tal-sh).'' \url{https://github.com/DmitryLyakh/TAL\_SH}.

\bibitem{TensorTranspose1}
D.~I. Lyakh, ``An efficient tensor transpose algorithm for multicore cpu, intel
  xeon phi, and nvidia tesla gpu,'' {\em Computer Physics Communications},
  vol.~189, pp.~84--91, 2015.

\bibitem{TensorTranspose2}
A.~Hynninen and D.~I. Lyakh, ``cutt: {A} high-performance tensor transpose
  library for {CUDA} compatible gpus,'' {\em CoRR}, vol.~abs/1705.01598, 2017.

\bibitem{DSVP}
D.~I. Lyakh, ``Domain-specific virtual processors as a portable programming and
  execution model for parallel computational workloads on modern heterogeneous
  high-performance computing architectures,'' {\em International Journal of
  Quantum Chemistry}, p.~e25926, 2019.

\bibitem{zhang2019alibaba}
F.~Zhang, C.~Huang, M.~Newman, J.~Cai, H.~Yu, Z.~Tian, B.~Yuan, H.~Xu, J.~Wu,
  X.~Gao, {\em et~al.}, ``Alibaba cloud quantum development kit: Large-scale
  classical simulation of quantum circuits,'' {\em arXiv preprint
  arXiv:1907.11217}, 2019.

\end{thebibliography}

\end{document}